\documentclass[conference]{IEEEtran}
\usepackage[T1]{fontenc}
\usepackage{epstopdf}
\usepackage{graphicx}
\usepackage{float}
\usepackage{nomencl}
\usepackage{balance} 
\usepackage{color,soul}
\usepackage{multirow}
\usepackage{makecell}
\usepackage{longtable}
\usepackage{hhline}
\usepackage{pifont}
\usepackage{tabularx}
\usepackage[inline,shortlabels]{enumitem}
\usepackage{xspace}
\usepackage[ruled,vlined,linesnumbered]{algorithm2e}
\SetKwInput{KwInput}{Input}
\SetKwInput{KwOutput}{Output}
\usepackage{url}
\usepackage[binary-units,per-mode=symbol,exponent-product=\cdot,
            list-units=single,list-final-separator={,}]{siunitx}
\DeclareSIUnit[number-unit-product = ]\pixel{p}
\DeclareSIUnit[number-unit-product = ]\dB{dB}
\usepackage{chemformula}
\usepackage{gensymb}
\usepackage{amsmath}
\usepackage{amssymb}
\usepackage{booktabs}
\usepackage{comment}
\usepackage{footnote}
\usepackage{tablefootnote}
\usepackage{caption}
\usepackage{subcaption} 
\usepackage{wrapfig}
\usepackage{xurl}
\usepackage{orcidlink}
\usepackage{hyperref}
\usepackage{hyperxmp}
\usepackage{footmisc}
\usepackage[table]{xcolor}
\hypersetup{pdfauthor=author}
\newcommand{\CL}{\texttt{ClusterLess}\xspace}

\newcommand{\etal}{et al.\xspace}

\newcommand{\ie}{i.e., }

\begin{document}
\title{ClusterLess: Deadline-Aware Serverless Workflow Orchestration on Federated Edge Clusters}
\author{\IEEEauthorblockN{Reza Farahani\IEEEauthorrefmark{1}, Mario Colosi\IEEEauthorrefmark{2}, Ilir Murturi\IEEEauthorrefmark{3}, Stefan Nastic\IEEEauthorrefmark{1}, Massimo Villari\IEEEauthorrefmark{2}, Schahram Dustdar\IEEEauthorrefmark{1}, Radu Prodan\IEEEauthorrefmark{4}
\IEEEauthorblockA{\IEEEauthorrefmark{1}Distributed Systems Group (DSG), TU Wien, Vienna, Austria}
}

\IEEEauthorblockA{\IEEEauthorrefmark{2} MIFT Department, University of Messina, Messina, Italy}

\IEEEauthorblockA{\IEEEauthorrefmark{3} Department of Mechatronics, University of Prishtina, Prishtina, Kosova}

\IEEEauthorblockA{\IEEEauthorrefmark{4} Department of Computer Science, University of Innsbruck, Innsbruck, Austria}
}
\maketitle
\begin{abstract}
The recent convergence of edge computing, serverless execution, and Kubernetes (K8s)-based container orchestration has enabled the processing of application workflows close to data sources. While effective within a single-edge cluster, existing schemes do not generalize to federated multi-edge environments, where multiple workflows execute concurrently under strict end-to-end (E2E) deadline constraints. 
This paper introduces \CL, a deadline-aware serverless workflow \emph{orchestration} method for federated multi-edge K8s clusters. \CL manages the E2E lifecycle of workflow execution, including dependency analysis, execution-mode selection, and resource-aware placement. To this end, it integrates structured \emph{intra-cluster} orchestration with a leader-selected, \emph{super-master}–driven \emph{inter-cluster coordination} layer, determining where and how each workflow function should be executed across the federated edge clusters. We implement \CL using OpenFaaS as the serverless execution substrate and Argo for workflow management, and deploy it on a realistic testbed of \emph{six} edge clusters comprising \num{64} heterogeneous edge nodes. Experimental results with concurrent serverless workflows, spanning \num{18} workload configurations across different input sizes and deadline classes, show that \CL reduces workflow completion time by up to \qty{40}{\percent}, increases deadline satisfaction from below \qty{50}{\percent} to over \qty{90}{\percent}, and confines deadline violations to single-digit seconds compared to four baseline methods.

\end{abstract}
\section{Introduction}
\label{sec:Introduction}
Recent industry analyses predict that over \qty{50}{\percent} of critical enterprise applications will run outside centralized public clouds or traditional data centers by \num{2027}, reflecting the accelerating shift toward distributed edge computing~\cite{gartner-2027}. In parallel, serverless computing has become one of the de facto cloud execution models, with production traces reporting billions of function invocations per day on commercial platforms~\cite{joosen2023does,farahani2024serverless}. To support this growth at the edge, Kubernetes (K8s), predominantly employed as the orchestration layer, provides container-based isolation, rapid scaling, and uniform resource management across heterogeneous edge clusters~\cite{mondal2022kubernetes,aslanpour2024faashouse}. While effective within a single edge cluster, these mechanisms do not readily generalize to federated multi-edge environments, where application workflows with strict end-to-end (E2E) deadlines must be executed concurrently across clusters with diverse compute capacities and network conditions~\cite{carrion2022kubernetes}. In such scenarios, cold starts, resource fragmentation, and inter-cluster communication delays can degrade performance, posing challenges to deadline-aware orchestration~\cite{farahani2025energyless, shafiei2022serverless, gill2024modern}.

Serverless workflows, typically expressed as directed acyclic graphs (DAGs), require fine-grained, function-level orchestration that respects dependency order, deadline constraints, and interference from concurrent executions. Misplacing an upstream function can propagate delays across the workflow.
for example, placing $f_1$ in a workflow \mbox{$f_1\rightarrow f_2\rightarrow f_3$} on a congested compute instance delays all downstream functions and can violate the E2E deadline despite available downstream resources. Although recent efforts have explored multi-cluster K8s~\cite{farahani2024heftless, michalke2024evaluating, bachar2023optimizing}, orchestration support for serverless workflows across multiple edge clusters remains limited. Moreover, existing centralized orchestration methods~\cite{poggiani2024live} often overlook workflow-level deadlines, dependency structures, and cross-cluster heterogeneity, limiting their effectiveness in real multi-edge environments with dynamic loads and asymmetric computing and networking conditions.

To address these challenges, we introduce \CL, a deadline-aware serverless workflow orchestration method for federated multi-edge K8s clusters. \CL orchestrates the execution of workflow functions by jointly considering dependency constraints, workflow deadlines, and interference from concurrent workflows. Each cluster runs a local master that performs \emph{intra-cluster orchestration}; for each function invocation, this master selects among four execution modes using native K8s and serverless mechanisms:
\begin{enumerate*}
\item\emph{warm execution} on edge nodes already hosting an active function instance and offering the lowest completion time;
\item \emph{warm scaling} when existing deployments are saturated but additional replicas remain deadline-feasible;
\item \emph{cold scaling} on suitable edge nodes when warm execution and autoscaling risk violating the workflow deadline;
\item \emph{offloading} when no local execution option can satisfy the workflow deadline.
\end{enumerate*}
In the offloading mode, control is transferred to \emph{inter-cluster orchestration}, where one cluster master is dynamically elected as a logically central \emph{super-master}. The super-master operates alongside its local master responsibilities, aggregates cluster-level state, and evaluates feasible execution placements across clusters based on communication delay, deployment overhead, queueing state, and execution time, minimizing E2E workflow latency while respecting deadline constraints. It is re-elected upon failure or overload, preserving orchestration continuity.

To our knowledge, \CL is the first orchestration method that jointly incorporates workflow-level deadline guarantees, DAG-aware function scheduling and execution, cross-cluster coordination, and resource heterogeneity into a unified orchestration model for K8s-based edge serverless environments. We implement \CL using OpenFaaS as the serverless execution substrate and Argo for workflow management, and deploy it on a real-world multi-cluster edge testbed comprising six K8s clusters. The testbed consists of \num{64} heterogeneous edge instances, including Jetson-class devices (Nano, Orin Nano, AGX), Raspberry Pis, and x86-based virtual machines. We evaluate \CL using concurrent serverless workflows with different deadline tightness and input sizes. Experiments show that \CL lowers completion time by up to \qty{40}{\percent} and raises deadline satisfaction from below \qty{50}{\percent} to over \qty{90}{\percent} under heterogeneous workloads.


\section{Related Work}
\label{sec:RW}
Multi-cluster K8s solutions interconnect and coordinate separate clusters, enabling seamless workload placement and migration (e.g., microservices) across heterogeneous infrastructures ranging from the edge to the cloud. 
Michalke~\etal~\cite{michalke2024evaluating} evaluated three multi-cluster connectivity solutions (Submariner, Clusternet, Skupper), demonstrating that inter-cluster communication overhead significantly affects latency and throughput for distributed microservices. Bachar \etal~\cite{bachar2023optimizing} introduced a multi-cluster optimized service-selection system for geo-distributed K8s deployments that employs a centralized broker with Domain Name System (DNS)-based routing to balance cost and latency, ignoring workflow dependencies and deadline constraints. Park~\etal~\cite{park2025heart} introduced a scheduler for K8s-based multi-replica services that profiles each replica’s performance and dynamically routes requests to the one with the lowest predicted E2E latency.
Early Cloud Native Computing Foundation (CNCF) initiatives such as KubeFed provided foundational support for propagating K8s resources across clusters, but focused primarily on resource replication rather than fine-grained orchestration. More recent systems, such as Karmada~\cite{karmada}, enable coarse-grained cross-cluster service discovery and failover, yet remain agnostic to serverless workflow orchestration and deadline guarantees.

\paragraph*{State-of-the-art limitations}
Existing systems address microservice placement and cross-cluster connectivity, yet lack deadline-aware orchestration for serverless workflows. By ignoring deadlines, function dependencies, and inter-cluster latency, they fall short in resource-limited edge environments.

Several serverless platforms like Knative and Fission leverage K8s primitives for elastic containerized function lifecycle management and resource allocation, but face challenges such as cold-start latency, resource oversubscription, and limited support for workflow-aware or dependency-sensitive orchestration~\cite{balla2020open}.
Lin and Glikson~\cite{lin2019mitigating} tackled Knative’s cold-start bottleneck by introducing a warm pool of pre-provisioned function containers, reducing response latency for sporadic or latency-sensitive invocations. Cvetković~\etal~\cite{cvetkovic2024dirigent} introduced a K8s-inspired cluster manager for function scheduling that employs a centralized function scheduler and a lightweight runtime for function invocation throughput, in contrast to traditional layered designs. 
Simion~\etal~\cite{simion2023towards} extended Knative with edge-aware offloading by leveraging latency estimates and location-aware placement, improving throughput for IoT workloads. López~\etal~\cite{lopez2020triggerflow} proposed Triggerflow, a trigger-based serverless workflow orchestrator for K8s/Knative, but it emphasizes event-driven control-flow extensibility rather than deadline-aware federated multi-edge orchestration.
Serenari~\etal~\cite{serenari2024greenwhisk} proposed GreenWhisk, an Apache OpenWhisk-based system evaluated on Raspberry Pi edge clusters, enabling carbon-aware energy-aware function placement. Poggiani~\etal~\cite{poggiani2024live} proposed live migration of multi-container K8s pods across clusters via container-level checkpointing to preserve state and reduce cold-start overheads, focusing on stateful relocation rather than workflow-level serverless orchestration.

\paragraph*{State-of-the-art limitations} Most existing systems improve per-function scheduling, but operate strictly within single clusters and offer no distributed orchestration across multiple K8s clusters. They also overlook functional dependencies, workflow-level deadlines, and cross-cluster computing and bandwidth constraints, limiting their use for concurrent serverless workflows in multi-cluster edge environments.

\begin{table}[t]
\centering
\renewcommand{\arraystretch}{1.1}
\setlength{\tabcolsep}{1.2pt}
\caption{\small{Related work comparison
(SL: Serverless; P: Pod; WF: Workflow; MC: Multi-cluster;
AT: Autoscaling; ND: New Deployment; DL: Deadline-aware;
Intra: Intra-cluster; Inter: Inter-cluster).}}
\label{tab:sota}
\fontsize{7pt}{7pt}\selectfont
\resizebox{\columnwidth}{!}{
\begin{tabular}{|c|c|c|c|c|c|c|c|c|c|c|}
\hline
\multirow{2}{*}{\textit{Work}}
& \multirow{2}{*}{\textit{SL}}
& \multicolumn{2}{c|}{\textit{Processing}}
& \multirow{2}{*}{\textit{MC}}
& \multirow{2}{*}{\textit{AT}}
& \multirow{2}{*}{\textit{ND}}
& \multirow{2}{*}{\textit{DL}}
& \multicolumn{2}{c|}{\textit{Orchestration}}
& \multirow{2}{*}{\textit{Evaluation infrastructure}} \\
\cline{3-4} \cline{9-10}
& & \textit{P} & \textit{WF}
& 
& 
& 
& 
& \textit{Intra} & \textit{Inter}
& \\
\hline
\cite{michalke2024evaluating}
& $\checkmark$ & $\checkmark$ & $\times$ & $\checkmark$
& $\times$ & $\times$
& $\times$
& $\times$ & $\checkmark$
& 2 k3s clusters, Knative.\\
\cite{bachar2023optimizing}
& $\times$ & $\checkmark$ & $\times$ & $\checkmark$
& $\times$ & $\times$
& $\times$
& $\times$ & $\checkmark$
& Simulation and 5 K8s clusters.\\
\cite{poggiani2024live}
& $\checkmark$   
& $\checkmark$        
& $\times$     
& $\checkmark$   
& $\times$       
& $\checkmark$  
& $\times$       
& $\times$       
& $\checkmark$  
& 2 K8s clusters on OpenStack.\\
\cite{park2025heart}
& $\times$          %
& $\checkmark$      
& $\times$          
& $\times$          
& $\times$          
& $\checkmark$      
& $\times$          
& $\checkmark$      
& $\times$         
& Single K8s cluster (5 nodes).\\
\cite{lin2019mitigating}
& $\checkmark$         
& $\checkmark$          
& $\times$              
& $\times$              
& $\checkmark$         
& $\checkmark$          
& $\times$              
& $\checkmark$          
& $\times$              
& Single K8s cluster, Knative.\\
\cite{cvetkovic2024dirigent}
& $\checkmark$       
& $\checkmark$       
& $\times$           
& $\times$           
& $\checkmark$       
& $\checkmark$        
& $\times$            
& $\checkmark$       
& $\times$           
& 93-node cluster, Knative.\\
\cite{simion2023towards}
& $\checkmark$          
& $\checkmark$         
& $\times$              
& $\checkmark$          
& $\times$              
& $\checkmark$          
& $\times$              
& $\times$          
& $\checkmark$          
& 4 RPis, x64 node, VM, Knative.\\
\cite{lopez2020triggerflow}
& $\checkmark$          
& $\checkmark$             
& $\checkmark$          
& $\times$              
& $\checkmark$          
& $\checkmark$          
& $\times$              
& $\checkmark$         
& $\times$              
& K8s cluster (5 nodes) and IBM cloud. \\
\cite{serenari2024greenwhisk}
& $\checkmark$             
& $\checkmark$                  
& $\times$                 
& $\times$                  
& $\checkmark$              
& $\checkmark$              
& $\times$                  
& $\checkmark$              
& $\times$                  
& 10 RPis and 10 servers, OpenWhisk.\\
\hline
\rowcolor{gray!50}
\textbf{\CL}
& $\checkmark$ & $\checkmark$ & $\checkmark$ & $\checkmark$
& $\checkmark$ & $\checkmark$
& $\checkmark$
& $\checkmark$ & $\checkmark$
& \num{64} Jetsons, RPis, VMs, OpenFaaS, Argo. \\
\hline
\end{tabular}%
}
\vspace{-5pt}
\end{table}
\section{Problem Formulation}\label{sec:model}
\subsection{Function and workflow model}
We consider a set of \emph{serverless workflows} $\mathcal{W}$ executed at different time intervals. Each workflow \mbox{$w\in\mathcal{W}$} is modeled as a DAG \mbox{$G_w = \left(F_w, E_w\right)$}, where $F_w$ denotes the set of workflow functions and \mbox{$E_w\subseteq F_w \times F_w$} denotes dependency edges, such that \mbox{$(g,f)\in E_w$} implies that function $f$ can start only after $g$ completes. Each workflow $w$ arrives at time $A_w$ and is subject to a strict end-to-end (E2E) deadline $D_w$, where delays in individual functions may propagate and result in workflow-level deadline violations. Each function \mbox{$f\in F_w$} is characterized by:
\begin{enumerate*}
    \item computational demands, given by $C^f$ (CPU-seconds) and $M^f$ (bytes of memory); and
    \item input and output data sizes \mbox{$(X_{\text{in}}^f, X_{\text{out}}^f)$} in \unit{\mega\byte}, which determine inter-function communication overhead.
\end{enumerate*}
Multiple workflows may execute concurrently and contend for compute, memory, and network bandwidth resources across federated edge clusters.
\subsection{Cluster model}
We consider a set of $N$ federated K8s edge clusters \mbox{$\mathcal{K}=\{K_1,\dots,K_N\}$} deployed at distinct locations. Each cluster \mbox{$K_n\in\mathcal{K}$} comprises:
\begin{enumerate*}
    \item a \emph{local master} $M_n$ responsible for \emph{intra-cluster} orchestration decisions, including execution-mode selection (warm execution, warm scaling, cold scaling, or offloading); and
    \item a set of \emph{worker nodes} $\mathcal{Z}_n=\{z_{n1},\dots,z_{n|\mathcal{Z}_n|}\}$ 
\end{enumerate*}
At time $t$, each worker \mbox{$z_{ni}\in\mathcal{Z}_n$} provides available CPU and memory capacities $C_{ni}(t)$ and $M_{ni}(t)$, and maintains a local execution queue, inducing a queuing delay $Q_{ni}(t)$ due to concurrent workloads. Let $\Gamma_{ni}(t)$ denote the number of active function instances on worker $z_{ni}$ at $t$, and let $R_{ni}$ be its maximum concurrency capacity. We define the normalized load of cluster $K_n$ at time $t$ as:
\begin{equation}
\mathrm{L}_n(t)=
\frac{1}{|\mathcal{Z}_n|}
\sum_{z_{ni}\in\mathcal{Z}_n}
\frac{\Gamma_{ni}(t)}{R_{ni}}.
\label{eq:1}
\end{equation}
which captures the average utilization of worker-level concurrency capacity and enables load comparison across clusters. 

While each local master independently performs intra-cluster orchestration, deadline-feasible execution cannot always be guaranteed locally due to resource contention, cold-start overheads, or bursty arrivals. To orchestrate \emph{inter-cluster} across clusters, the system maintains a \emph{logically centralized} coordinator, referred to as the \emph{super-master}, elected among the local masters $\{M_1,\dots,M_N\}$. We model the super-master as a time-indexed selection function \mbox{$SM(t_e)\in\mathcal{K}\cup\{\emptyset\}$}, where \mbox{$t_e=e\cdot\Delta T$} denotes a discrete control epoch and \mbox{$SM(t_e)=\emptyset$} 
represents a transient state in which no eligible super-master is available (e.g., due to failures or overload). Each local master $M_n$ periodically emits heartbeat messages to signal
its availability. Let $t^{HB}_n$ denote the most recent heartbeat received from $M_n$.
Cluster $K_n$ is considered \emph{alive} at epoch $t_e$ if:
\begin{equation}
\mathrm{Alive}_n(t_e)=
\begin{cases}
1, & t_e - t^{HB}_n \le T_{\mathrm{fail}};\\
0, & \text{otherwise}.
\end{cases}
\label{eq:2}
\end{equation}
where $T_{\mathrm{fail}}$ is the failure-detection timeout. Thus, only clusters that satisfy Eq.~\ref{eq:3} are eligible to act as super-master:
\begin{equation}
\mathrm{Alive}_{n}(t_e)=1
\quad\land\quad
\mathrm{L}_{n}(t_e)\le l,
\label{eq:3}
\end{equation}
where $l$ shows the acceptable super-master overhead. 
\subsection{Intra-cluster model}
When a function \mbox{$f\in F_w$} becomes \emph{ready} for execution (\ie all predecessor functions in $E_w$ have completed), the local master $M_n$ of the cluster $K_n$ to which the request is initially
submitted performs an \emph{intra-cluster orchestration} decision. For each ready function $f$, the local master selects an execution mode $em^{f}$ from the following set:
\subsubsection{Warm execution} serves $f$  by an already active container on worker $z_{ni}$, incurring only queuing delay $Q_{ni}(t)$ and execution time $\mathcal{T}^f_{ni}$.
\subsubsection{Warm scaling} spawns an additional replica of $f$ on $z_{ni}$ using existing images, incurring an autoscaling delay $WS^f_{ni}$ before execution.
\subsubsection{Cold scaling} deploys $f$ from scratch (image pull, initialization, and container setup) on $z_{ni}$, incurring a cold-start overhead $CS^f_{ni}$ prior to execution.
\subsubsection{Offloading} forwards the execution request to the super-master $SM(t)$ for placement on a remote cluster.

If $f$ is assigned locally to worker $z_{ni}$ under execution mode $em^{f}$, its \emph{serving time} $\chi^f_n$ is given by:
\begin{equation}
\chi^f_n =
\begin{cases}
Q_{ni}(t) + \mathcal{T}^f_{ni}, & \hspace{-.3cm}em^f = \text{warm execution};\\[2pt]
Q_{ni}(t) + WS^f_{ni} + \mathcal{T}^f_{ni}, & em^f = \text{warm scaling};\\[2pt]
Q_{ni}(t) + CS^f_{ni} + \mathcal{T}^f_{ni}, & em^f = \text{cold scaling}.
\end{cases}
\label{eq:4}
\end{equation}
For dependent functions \mbox{$g\rightarrow f\in E_w$}, the data-transfer delay $\theta^{gf}$ is:
\begin{equation}
\theta^{gf}_n =
\begin{cases}
\dfrac{X_{\text{out}}^{g}}{b_{n,i,j}(t)}, & g \text{ and } f \text{ run on } z_{ni}\neq z_{nj} \text{ in } K_n;\\[4pt]
0, & g \text{ and } f \text{ run on the same worker } z_{ni}.
\end{cases}
\label{eq:5}
\end{equation}
where $b_{n,i,j}(t)$ is the available intra-cluster bandwidth. We define $\mathcal{S}^f_n$ as the start time of $f$ conditional on executing in $K_n$ defined in Eqs.~(\ref{eq:6}):
\begin{equation}
\mathcal{S}^f_n =
\begin{cases}
A_w, & \nexists\, g \rightarrow f \in E_w;\\[2pt]
\max\limits_{g \rightarrow f \in E_w}\Bigl\{\mathcal{F}^g_n + \theta^{gf}_n\Bigr\}, &
\exists\, g \rightarrow f \in E_w.
\end{cases}
\label{eq:6}
\end{equation}
where $\mathcal{F}^g_n$ denotes the completion time of predecessor function $g$ when executed in $K_n$, computed by Eq.~(\ref{eq:7}).
\begin{equation}
\mathcal{F}^g_n = \mathcal{S}^g_n + \chi^g_n.
\label{eq:7}
\end{equation}
\subsection{Inter-cluster model}
If the local master $M_n$ cannot place a ready function $f\in F_w$ within $K_n$ such that the workflow deadline constraint can still be satisfied, it offloads $f$ to the super-master $SM(t_e)$ at control epoch $t_e$. For each candidate cluster $K_{n'}\in\mathcal{K}\setminus\{K_n\}$, $SM(t_e)$ calculates the function completion time of $f$ as:
\begin{equation}
\mathcal{F}^{f}_{n'} =
\max\!\left(\mathcal{S}^f_{n'},\, t_e\right)
+ \delta_{nn'} + \chi^{f}_{n'}.
\label{eq:9}
\end{equation}
\begin{figure*}[!t]
\centering
\includegraphics[width=1\textwidth]{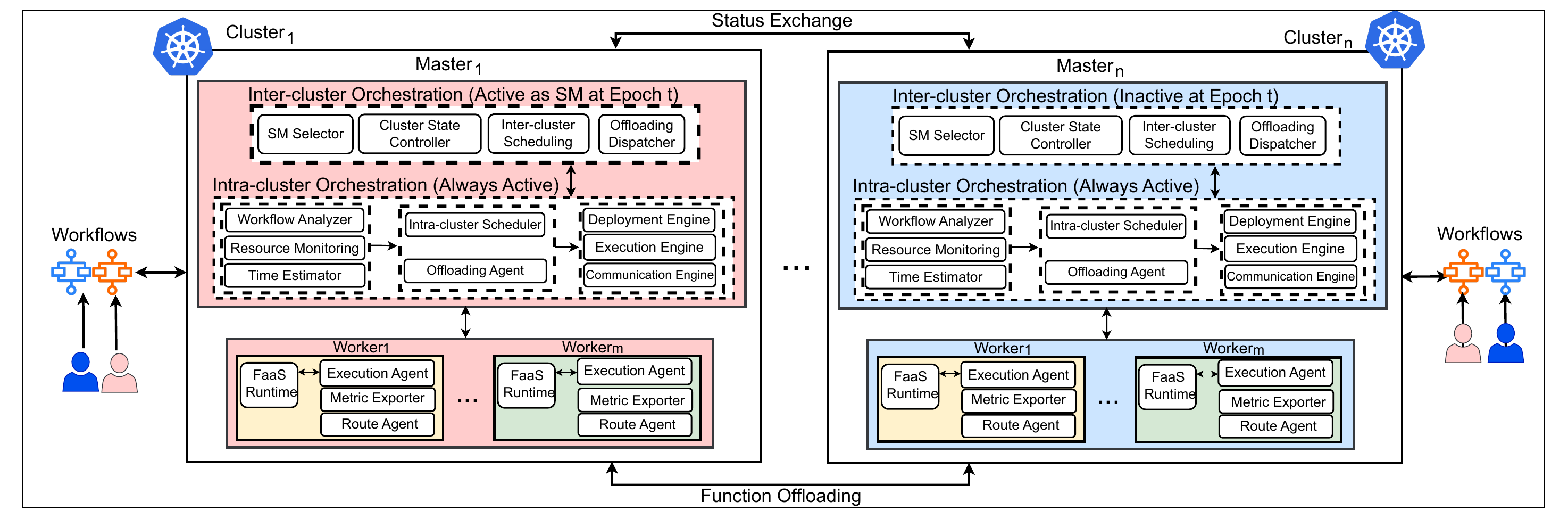}
\caption{\small{\CL system architecture.}} 
\label{fig:arch}
\vspace{-8pt}
\end{figure*}
where $\mathcal{S}^f_{n'}$ is computed via Eq.~(\ref{eq:6}) based on the remote cluster state, $\delta_{nn'}$ is the inter-cluster transfer delay from cluster $K_n$ to $K_{n'}$  and $\chi^{f}_{n'}$ is the serving time of $f$ in cluster $K_{n'}$, computed using the same execution modes as in Eq.~(\ref{eq:4}) based on the remote cluster state. The super-master considers the set of feasible target clusters only when the cluster satisfies the deadline-feasibility constraint:
\begin{equation}
\mathcal{K}^f =
\left\{\, K_{n'} \in \mathcal{K}\setminus\{K_n\}
\ \middle|\ 
\mathcal{F}^{f}_{n'} \le A_w + D_w \right\}.
\label{eq:10}
\end{equation}

If \mbox{$\mathcal{K}^f\neq\varnothing$}, $SM(t_e)$ selects the destination cluster $K_{n^*}$ as:
\begin{equation}
K_{n^*} = \arg\min_{K_{n'}\in\mathcal{K}^f} \mathcal{F}^{f}_{n'}.
\label{eq:11}
\end{equation}
The selection of $K_{n^*}$ constitutes a binding inter-cluster placement decision, and function $f$ is dispatched to cluster $K_{n^*}$ for execution. If \mbox{$\mathcal{K}^f=\varnothing$}, no deadline-feasible inter-cluster placement exists; function $f$ is declared infeasible, and workflow $w$ is marked as deadline-violated. In addition, when multiple offloading requests are pending at $SM(t_e)$, it processes them according to an earliest-deadline-first policy.
\subsection{Resource feasibility model}
A placement decision for function $f$ on worker $z_{ni}$ at time $t$ is \emph{resource-feasible} only if sufficient compute and memory resources are available on the worker node:
\begin{equation}
C^f \le C_{ni}(t),
\qquad
M^f \le M_{ni}(t),
\label{eq:12}
\end{equation}

In addition, data transfers induced by workflow dependencies must be \emph{bandwidth-feasible}. For a dependency \mbox{$g\rightarrow f\in E_w$} executed on workers $z_{ni}$ and $z_{n j}$ within the same cluster $K_n$, the required data transfer is feasible only if:
\begin{equation}
X_{\text{out}}^{g} \le b_{n,i,j}(t),
\label{eq:13}
\end{equation}
where $b_{n,i,j}(t)$ denotes the currently available intra-cluster bandwidth between the two workers at time $t$. For inter-cluster execution, when $g$ is executed in cluster $K_n$ and $f$ is offloaded to cluster $K_{n'}$, the dependency transfer is feasible only if:
\begin{equation}
X_{\text{out}}^{g} \le b_{n,n'}(t),
\label{eq:14}
\end{equation}
where $b_{n,n'}(t)$ denotes the available inter-cluster bandwidth between
clusters $K_n$ and $K_{n'}$.
\subsection{Completion-time model}
For each function $f\in F_w$, the orchestration process (intra- or inter-cluster execution) induces a unique execution cluster. Accordingly, the effective completion time of $f$ is:
\begin{equation}
\mathcal{F}^f =
\begin{cases}
\mathcal{F}^f_n, &  f \text{ is executed in } K_n;\\
\mathcal{F}^f_{n^*}, & f \text{ is offloaded and executed in } K_{n^*}.
\end{cases}
\label{eq:15}
\end{equation}

The end-to-end (E2E) completion time of workflow $w$ is defined in Eq.~(\ref{eq:16}) and is considered \emph{deadline-feasible} if it satisfies the workflow deadline.
\begin{equation}
C_w = \max_{f\in F_w} \mathcal{F}^f\le A_w + D_w.
\label{eq:16}
\end{equation}

\section{\CL Architecture}\label{sec:Arch}
Fig.~\ref{fig:arch} depicts the \CL architecture spanning $N$ federated K8s-based edge clusters. Each cluster consists of a \emph{local master} responsible for control-plane decisions and a set of \emph{worker nodes} that execute serverless functions.
\subsubsection{Local master nodes} support two logically distinct orchestration paths, corresponding to intra- and inter-cluster decision-making, which jointly enable deadline-aware workflow execution across the federation.
\paragraph{Intra-cluster orchestration} is \emph{always active} on every master and realizes local workflow execution decisions within a cluster. Upon workflow submission, the \emph{workflow analyzer} parses the workflow structure and dependency relations, while the \emph{resource monitoring} module continuously tracks worker-level compute, memory, queuing, and bandwidth conditions. Based on this information, the \emph{time estimator} derives execution-time estimates for different execution modes (warm execution, warm scaling, cold scaling). The \emph{intra-cluster scheduler} then determines a feasible worker and execution mode for each ready function. When no such placement exists locally, the \emph{offloading agent} escalates the function to the inter-cluster orchestration path. The selected decisions are enforced through the \emph{deployment}, \emph{execution}, and \emph{communication} engines, which collectively realize the selected placement, execution mode, and dependency-aware data transfers on worker nodes.
\paragraph{Inter-cluster orchestration} is deployed on every master but \emph{remains inactive} unless the master is designated as the super-master. When active, it enables orchestration by aggregating cluster-level state and arbitrating offloaded functions. The \emph{SM selector} maintains cluster liveness and load information, while the \emph{cluster state controller} constructs a global view of the federation. Using this information, the \emph{inter-cluster scheduler} evaluates offloaded functions across clusters considering their execution feasibility and deadline urgency, and the \emph{offloading dispatcher} communicates the resulting placement decisions back to the destination masters. Inter-cluster orchestration relies on two logical communication paths: a \emph{status exchange} disseminates load and liveness information among masters, and a \emph{function offloading} transfers execution requests and placement decisions across clusters.
\subsubsection{Worker nodes} implement the data-plane components required to execute functions delegated by the local master. Each worker hosts a lightweight serverless \emph{FaaS runtime} and a minimal set of agents, including an \emph{execution agent} for triggering function execution, a \emph{metric exporter} for reporting execution and resource statistics, and a \emph{route agent} for managing dependency-aware data transfers between functions on the appropriate intra- or inter-cluster links.

\section{\CL Decision-Marking Algorithms}\label{sec:Alg}
\subsection{Super-master maintenance}\label{sec:SM}
Alg.~\ref{alg:SM_selection} presents the epoch-based super-master maintenance procedure, ensuring that \CL operates under a responsive, load-aware, and fault-tolerant coordinator for inter-cluster orchestration. The algorithm takes as input the set of clusters $\mathcal{K}$, the control epoch length $\Delta T$, the heartbeat failure timeout $T_{\mathrm{fail}}$, and the admissible coordination-load threshold $l$, and publishes the super-master $SM(t_e)$ at each epoch $t_e$. At system startup (\mbox{$e=0$}), each cluster initializes its heartbeat timestamp and computes its normalized load $\mathrm{L}_n(t_0)$ based on worker concurrency (lines~2--4). The cluster with the minimum load is deterministically selected as the initial super-master (line~5), yielding a lightweight coordinator at bootstrap. At each subsequent epoch \mbox{$t_e=e\cdot\Delta T$}, clusters update their load estimates if a heartbeat is received during $(t_{e-1},t_e]$; otherwise, the previous value is retained to avoid oscillations under transient reporting delays (lines~6--12). 

Cluster liveness is then evaluated using the timeout condition in Eq.~(\ref{eq:2}) (lines~13--14). The current super-master is validated against the eligibility constraint in Eq.~(\ref{eq:3}); it is invalidated if it is unavailable, not alive, or exceeds the admissible load threshold (line~15). If invalid, the algorithm deterministically re-selects the alive and load-eligible cluster with the minimum normalized load (lines~16--17); if no such cluster exists, no super-master is assigned for the epoch (line~19), deferring inter-cluster coordination to subsequent epochs. Otherwise, the previous super-master is retained unchanged (line~21). At the end of each epoch, the resulting $SM(t_e)$, either newly selected, retained, or empty, is published to the inter-cluster orchestration layer (line~22), enabling continuous, stable, and epoch-consistent decisions.
\begin{algorithm}[!t]
{\fontsize{7pt}{7pt}\selectfont
\caption{\small{Super-master maintenance.}}
\label{alg:SM_selection}
\KwInput{$\mathcal{K}=\{K_1,\dots,K_N\}$, $\Delta T$, $T_{\mathrm{fail}}$, $l$}
\KwOutput{$SM(t_e)$}
$e \leftarrow 0$\quad $t_0 \leftarrow 0$ \\
\ForAll{$K_n \in \mathcal{K}$}{
  $t^{HB}_n \leftarrow t_0$ \\
  $\mathrm{L}_n(t_0) \leftarrow
  \dfrac{1}{|\mathcal{Z}_n|}
  \sum\limits_{z_{ni}\in\mathcal{Z}_n}
  \dfrac{\Gamma_{ni}(t_0)}{R_{ni}}$
}
$SM(t_0) \leftarrow \arg\min\limits_{K_n\in\mathcal{K}} \mathrm{L}_n(t_0)$ \\
\For{$e \leftarrow 1$ \KwTo $\infty$}{
  $t_e \leftarrow e\cdot\Delta T$ \\
  \ForAll{$K_n \in \mathcal{K}$}{
    \If{$t^{HB}_n > t_{e-1}$}{
      $\mathrm{L}_n(t_e) \leftarrow
      \dfrac{1}{|\mathcal{Z}_n|}
      \sum\limits_{z_{ni}\in\mathcal{Z}_n}
      \dfrac{\Gamma_{ni}(t_e)}{R_{ni}}$
    }
    \Else{
      $\mathrm{L}_n(t_e) \leftarrow \mathrm{L}_n(t_{e-1})$
    }
  }
  \ForAll{$K_n \in \mathcal{K}$}{
    $\mathrm{Alive}_n(t_e) \leftarrow
    \mathbb{I}\!\left[t_e - t^{HB}_n \le T_{\mathrm{fail}}\right]$
  }
  \If{$SM(t_{e-1})=\emptyset \ \lor\
      (\mathrm{Alive}_{SM(t_{e-1})}(t_e)=0 \ \lor\
      \mathrm{L}_{SM(t_{e-1})}(t_e)>l)$}{
    \If{$\exists\,K_n \in \mathcal{K}:\ \mathrm{Alive}_n(t_e)=1 \ \land\ \mathrm{L}_n(t_e)\le l$}{
      $SM(t_e) \leftarrow
      \arg\min\limits_{K_n\in\mathcal{K}:\mathrm{Alive}_n(t_e)=1\land \mathrm{L}_n(t_e)\le l}
      \mathrm{L}_n(t_e)$
    }
    \Else{
      $SM(t_e) \leftarrow \emptyset$
    }
  }
  \Else{
    $SM(t_e) \leftarrow SM(t_{e-1})$
  }
  \textsc{Publish}($SM(t_e)$)
}}
\end{algorithm}
\subsection{Intra-cluster orchestration}\label{sec:intra}
Alg.~\ref{alg:intra} presents the intra-cluster orchestration procedure executed by the local master $M_n$ when a workflow $w$ is submitted to cluster $K_n$ at time $t$. The algorithm takes as input the workflow DAG $G_w$, its arrival time $A_w$, deadline $D_w$, and the target cluster $K_n$, and outputs the local orchestration plan $\mathit{LOrch}$, specifying for each function $f\in F_w$ its execution mode $em^f$, local placement $z^f$ (if any), and the corresponding start and completion times $(\mathcal{S}^f_n,\mathcal{F}^f_n)$. At initialization (line~1), all functions are marked as local ($\textit{ext}=0$), assigned the default execution mode \emph{warm execution}, and left unscheduled with infinite completion time. The algorithm then iteratively refines $\mathit{LOrch}$ while the workflow-level completion time exceeds the deadline, i.e., $\textsc{WFinish}(\mathit{LOrch}) > A_w + D_w$ (line~2), where $\textsc{WFinish}()$ implements Eq.~(\ref{eq:16}), ensuring correctness for arbitrary DAG structures, including parallel terminal branches. In each iteration, $\textsc{IdentifyBottleneck}(\mathit{LOrch})$ is invoked (line~3) to extract three components: a committed partial plan $\mathit{LOrch}_c$, a bottleneck function $f^{bn}$ selected among non-offloaded functions with the largest current completion time, and a pending set $P$ containing all other functions whose execution may be affected by changes to $f^{bn}$. To ensure progress, functions that have already been evaluated under all execution modes without yielding an improvement are excluded from bottleneck selection in the current refinement cycle; however, a function may be selected again as a bottleneck if updates to predecessor functions reduce its effective completion time.

The algorithm initializes a control flag $success$ (line~5) to track whether an improving execution-mode update for the bottleneck function can be found. While no improvement is achieved and the bottleneck has not been escalated to offloading (lines~6--8), the execution mode of $f^{bn}$ is deterministically advanced using $\textsc{NextMode}()$ following the ordered policy \emph{warm execution}$\rightarrow$\emph{warm scaling}$\rightarrow$\emph{cold scaling}$\rightarrow$\emph{offloading}. For each mode, $\textsc{ApplyMode}()$ evaluates feasibility using the serving-time model in Eq.~(\ref{eq:4}), the dependency constraints in Eqs.~(\ref{eq:6})--(\ref{eq:7}), and the resource and bandwidth feasibility conditions in Eqs.~(\ref{eq:12})--(\ref{eq:14}). The flag $success$ is set to $1$ only if a feasible and improving update is obtained; offloading succeeds only upon acknowledgement from the super-master. If an improvement is found, the updated bottleneck decision is committed to $\mathit{LOrch}$ (lines~9--10). Otherwise, if all execution modes fail to yield a feasible update, the pending set $P$ is restored unchanged (lines~11--13), and the algorithm proceeds to the next iteration without altering prior decisions.
\begin{algorithm}[!t]
\fontsize{7pt}{7pt}\selectfont
\caption{\small{Intra-cluster orchestration by $M_n$ at time $t$.}}
\label{alg:intra}
\KwInput{$w$ with $G_w=(F_w,E_w)$, $A_w$, $D_w$, $K_n$, $t$}
\KwOutput{$\mathit{LOrch}$}
$\mathit{LOrch} \gets \{(w,f,\textit{ext}=0,z^f=\emptyset,em^f=\text{warm exec},
\mathcal{S}^f_n=\infty,\mathcal{F}^f_n=\infty)\mid f\in F_w\}$ \\
\While{$\textsc{WFinish}(\mathit{LOrch}) > A_w + D_w$}{
  $(\mathit{LOrch}_c,f^{bn},P) \gets \textsc{IdentifyBottleneck}(\mathit{LOrch})$ \\
  $\mathit{LOrch} \gets \mathit{LOrch}_c$ \\
  $success \gets 0$ \\
  \While{$(success=0)\ \land\ (em^{f^{bn}} \neq \text{offloading})$}{
    $em^{f^{bn}} \gets \textsc{NextMode}(em^{f^{bn}})$ \\
    $(success,fs^{bn},P) \gets
    \textsc{ApplyMode}(f^{bn},em^{f^{bn}},\mathit{LOrch},P,K_n,t)$
  }

  \eIf{$success=1$}{
    $\mathit{LOrch} \gets \mathit{LOrch} \cup \{fs^{bn}\}$
  }{
    $\mathit{LOrch} \gets \mathit{LOrch} \cup P$\\
    \textbf{continue}
  }
  $\mathit{LOrch}_p \gets \varnothing$ \\
  \ForAll{$fs \in P$}{
    $(w,f,\textit{ext},z^f,em^f,\mathcal{S}^f_n,\mathcal{F}^f_n) \gets fs$ \\
    \If{$\textit{ext}=1$}{\textbf{continue}}
    $\mathcal{I}^{f}_{\text{avail}} \gets \textsc{GetAvail}(f,em^f,K_n,t)$
    $fs' \gets \textsc{FuncOrch}(\mathit{LOrch}\cup\mathit{LOrch}_p,\ fs,\ \mathcal{I}^{f}_{\text{avail}},\ K_n,\ t)$ \\
    $\mathit{LOrch}_p \gets \mathit{LOrch}_p \cup \{fs'\}$
  }
  $\mathit{LOrch} \gets \mathit{LOrch} \cup \mathit{LOrch}_p$
}
\Return{$\mathit{LOrch}$}
\end{algorithm}

After committing the bottleneck update, the algorithm orchestrates the remaining pending functions (lines~15--20). For each function \mbox{$fs\in P$} that is not marked external, $M_n$ retrieves the set of available execution instances via $\textsc{GetAvail}()$ (line~19), restricting evaluation to runnable pods or scalable targets. The function-level orchestration module $\textsc{FuncOrch}()$ then selects a resource- and bandwidth-feasible placement and computes $(\mathcal{S}^f_n,\mathcal{F}^f_n)$ according to Eqs.~(\ref{eq:4})--(\ref{eq:7}) and Eqs.~(\ref{eq:12})--(\ref{eq:14}) (line~20). These decisions are merged into $\mathit{LOrch}$ (lines~20--21), completing one refinement step. The process terminates once the workflow deadline is satisfied, and the final plan $\mathit{LOrch}$ is returned (line~22).
\subsection{Inter-cluster orchestration}\label{sec:inter}
Alg.~\ref{alg:inter} presents the inter-cluster orchestration procedure executed by the super-master $SM(t_e)$ at control epoch $t_e$ for handling the set of offloaded functions $\mathit{Offload}(t_e)$. The algorithm takes as input the set of pending offload requests $\mathit{Offload}(t_e)$ and the federated cluster set $\mathcal{K}$, and outputs the global orchestration plan $\mathit{GOrch}(t_e)$, specifying one destination cluster per offloaded function. At the beginning of the epoch, all offload requests are ordered by earliest-deadline-first (EDF) using the workflow deadline $A_w+D_w$ (line~1), prioritizing functions on critical deadline paths. The global decision set $\mathit{GOrch}(t_e)$ is then initialized to empty (line~2). The super-master evaluates all candidate destination clusters for each offload request in EDF order (lines~3--4) for each candidate,  computes the dependency-aware earliest start time $\mathcal{S}^f_{n'}$ based on predecessor completion times and inter-cluster transfer delays (lines~5--8), followed by the corresponding remote completion time $\mathcal{F}^f_{n'}$ at epoch $t_e$ (line~9).

The algorithm then constructs the feasible-improving set $\mathcal{K}^f$ (line~10), containing only those clusters that both satisfy the workflow deadline constraint and strictly improve upon the best local completion time $\mathcal{F}^f_n$. If no such cluster exists, the function is retained at its origin cluster $K_n$ and recorded accordingly in $\mathit{GOrch}(t_e)$ (lines~11--12). Otherwise, the destination cluster that minimizes $\mathcal{F}^f_{n'}$ is selected (lines~14--15). This process is repeated for all offload functions, and $\mathit{GOrch}(t_e)$ is returned at the end of the epoch (line~16).
\begin{algorithm}[!t]
\fontsize{7pt}{7pt}\selectfont
\caption{\small{Inter-cluster orchestration by $SM(t_e)$.}}
\label{alg:inter}
\KwInput{$t_e$, \mbox{$\mathit{Offload}(t_e)=\{(w,f,K_n,\mathcal{F}^f_n)\}$}, $\mathcal{K}$}
\KwOutput{$\mathit{GOrch}(t_e)$}
$\mathit{Offload}(t_e)\leftarrow
\operatorname{argsort}_{(w,f,K_n,\mathcal{F}^f_n)\in\mathit{Offload}(t_e)}(A_w+D_w)$ \\
$\mathit{GOrch}(t_e)\leftarrow \varnothing$ \\

\ForAll{$(w,f,K_n,\mathcal{F}^f_n)\in\mathit{Offload}(t_e)$}{
  \ForAll{$K_{n'}\in \mathcal{K}\setminus\{K_n\}$}{
    \uIf{$\nexists\, g\!\to\! f\in E_w$}{
      $\mathcal{S}^f_{n'} \leftarrow A_w$
    }
    \Else{
      $\mathcal{S}^f_{n'} \leftarrow
      \max\limits_{g\to f\in E_w}\{\mathcal{F}^g_{n'}+\theta^{gf}_{n'}\}$
    }
    $\mathcal{F}^{f}_{n'} \leftarrow
    \max\{\mathcal{S}^f_{n'},\, t_e\} + \delta_{nn'} + \chi^{f}_{n'}$
  }

  $\mathcal{K}^f \leftarrow
  \left\{\, K_{n'} \in \mathcal{K}\setminus\{K_n\}
  \ \middle|\ 
  \mathcal{(F}^{f}_{n'} \le A_w+D_w) \ \land\ (\mathcal{F}^{f}_{n'} < \mathcal{F}^{f}_{n})
  \right\}$

  \uIf{$\mathcal{K}^f=\varnothing$}{
    $\mathit{GOrch}(t_e)\leftarrow \mathit{GOrch}(t_e)\cup\{(w,f,K_n)\}$
  }
  \Else{
    $K_{n^*}\leftarrow \arg\min\limits_{K_{n'}\in\mathcal{K}^f}\mathcal{F}^{f}_{n'}$ \\
    $\mathit{GOrch}(t_e)\leftarrow \mathit{GOrch}(t_e)\cup\{(w,f,K_{n^*})\}$
  }
}
\Return{$\mathit{GOrch}(t_e)$}
\end{algorithm}
\subsection{Time complexity}\label{sec:complexity}
Alg.~\ref{alg:SM_selection} performs a linear scan over clusters once per epoch, yielding $O(|\mathcal{K}|)$. Alg.~\ref{alg:intra} updates the local plan by resolving a single bottleneck per iteration across a bounded mode set, giving $O(|F_w|)$ per workflow. Alg.~\ref{alg:inter} sorts offloaded functions by EDF and evaluates all destination clusters, resulting in $O(|\mathit{Offload}|\log|\mathit{Offload}| + |\mathit{Offload}|\cdot|\mathcal{K}|)$ per epoch. 

\section{Evaluation Setup}
\label{sec:eval_set}
\subsection{Multi-cluster edge testbed}
We evaluate \CL on a realistic edge testbed comprising six K8s clusters, spanning \num{50} physical edge devices and \num{14} edge \texttt{KVM}-virtualized machines. All clusters run \texttt{Kubernetes~1.32} with \texttt{containerd~1.7} and are interconnected via \texttt{Submariner} 
(Globalnet mode), enabling transparent cross-cluster pod addressing, service discovery, and function offloading. Table~\ref{tab:testbed} summarizes the hardware composition, while Table~\ref{tab:clusters} details the per-cluster distribution of workers.
We employ \texttt{Argo Workflows 3.6} 
to orchestrate workflows, where each workflow step corresponds to a containerized \texttt{OpenFaaS} 
function invoked through HTTPS templates. \texttt{OpenFaaS} acts as the FaaS execution substrate for \CL, providing per-node function runtimes and exposing warm execution, warm scaling, cold scaling modes. All function images are stored in a \texttt{Harbor} 
registry to ensure consistent versioning and low-latency pulls. Workflow inputs, intermediate artifacts, and final outputs are stored in an S3-compatible \texttt{MinIO} 
backend used by both \texttt{Argo} and \texttt{OpenFaaS}, guaranteeing storage-consistent, cross-cluster execution and seamless offloading of data-dependent functions.

We emulate time-varying network conditions using six independent \texttt{4G LTE} bandwidth traces~\cite{raca2018beyond}, enforced via Linux traffic control with \texttt{wondershaper}. Each cluster is assigned a distinct trace, and worker-level traces are phase-shifted to avoid synchronized bandwidth fluctuations. The imposed bandwidth limits uniformly affect both inter-cluster control traffic (e.g., heartbeats and state exchange) and data-plane transfers, ensuring consistent and realistic network dynamics.
We collect system telemetry using \texttt{Prometheus}, \texttt{cAdvisor}, and scripts, providing fine-grained metrics on resource utilization and function lifecycle that \CL uses for monitoring during experiments.
\begin{table}[!t]
\centering
\caption{\small{Hardware composition of the six-cluster edge testbed.}}
\label{tab:testbed}
\setlength{\tabcolsep}{1pt}
\fontsize{6pt}{6pt}\selectfont
\begin{tabular}{|l|l|c|c|c|c|}
\hline
\textit{Node type} & \textit{Node class} & \textit{CPU cores} & \textit{RAM} & \textit{GPU cores} & \textit{Count} \\ \hline
\multirow{3}{*}{\emph{Intel VMs}}
  & XLarge VM      & 12 & 32 GB & -- & 2 \\ \cline{2-6}
  & Large VM      & 8  & 32 GB & -- & 2 \\ \cline{2-6}
  & Medium VM     & 4  & 24 GB & -- & 6 \\ \cline{2-6}
  & Small VM      & 2  & 16 GB & -- & 4 \\ \hline
\multirow{4}{*}{\emph{RPis}}
  & RPi 4         & 4  & 4 GB  & -- & 25 \\ \cline{2-6}
  & RPi 4B        & 4  & 4 GB  & -- & 4  \\ \cline{2-6}
  & RPi 3B+       & 2  & 1 GB  & -- & 9  \\ \hline
\multirow{3}{*}{\emph{Jetsons}}
  & Jetson Nano (JN)      & 4  & 4 GB  & 128 & 6 \\ \cline{2-6}
  & Jetson Orin Nano (JON)  & 6  & 8 GB  & 1024  & 2\\ \cline{2-6}
  & Jetson Orin AGX (JOA)   & 12 & 64 GB & 2048  & 2 \\ \hline
\emph{AMD server} 
  & Physical server  & 24 & 32 GB & 1536 & 2 \\ \hline
\end{tabular}
\end{table}
\begin{table}[!t]
\centering
\caption{\small{Summary of the six-cluster edge testbed.}}
\label{tab:clusters}
\setlength{\tabcolsep}{1pt}
\fontsize{6pt}{6pt}\selectfont
\begin{tabular}{|c|c|c|c|cccc|ccc|ccc|}
\hline
\multirow{3}{*}{\textit{Cluster\#}} 
& \multirow{3}{*}{\textit{Master}} 
& \multirow{3}{*}{\makecell{\textit{Total}\\\textit{CPU}}}
& \multirow{3}{*}{\makecell{\textit{Total}\\\textit{Memory (GB)}}}
& \multicolumn{10}{c|}{\textit{Workers}} \\ \cline{5-14}
& & & 
& \multicolumn{4}{c|}{\textit{VMs}} 
& \multicolumn{3}{c|}{\textit{RPis}} 
& \multicolumn{3}{c|}{\textit{Jetsons}} \\ \cline{5-14}
& & &
& \textit{XL} & \textit{L} & \textit{M} & \textit{S} 
& \textit{4B} & \textit{4} & \textit{3B+} 
& \textit{JN} & \textit{JON} & \textit{JOA} \\ \hline

$C_1$ & AMD server  & 82 & 210
& -- & -- & 2 & 2
& 1  & 3  & 2
& 2  & 1  & 1 \\ \hline

$C_2$ & AMD server  & 76 & 202
& -- & -- & 2 & 2
& 1  & 3  & 2
& 2  & -- & 1 \\ \hline

$C_3$ & Large VM  & 54 & 98
& -- & -- & 1 & --
& 1  & 6  & 2
& 1  & 1  & -- \\ \hline

$C_4$ & Large VM  & 36 & 80
& -- & -- & 1 & --
& 1  & 4  & --
& 1  & -- & -- \\ \hline

$C_5$ & XLarge VM  & 34 & 51
& -- & -- & -- & --
& -- & 4  & 3
& -- & -- & -- \\ \hline

$C_6$ & XLarge VM  & 28 & 48
& -- & -- & -- & --
& -- & 4  & --
& -- & -- & -- \\ \hline
\end{tabular}
\vspace{-10pt}
\end{table}
\subsection{Case-study serverless workflows}
We reimplemented two serverless applications from prior open-source systems: a \textit{text-to-speech censoring (T2SC)}~\cite{EiGrEyHeKo2020} and a \textit{regression-model training} (RT)~\cite{RTW}. We containerized both workflows and made them compatible with \texttt{OpenFaaS} and \texttt{Argo}, enabling automated DAG execution, artifact propagation, and seamless integration with \CL intra- and inter-cluster orchestration schemes. Each workflow is evaluated under three input sizes (\textit{small}, \textit{medium}, \textit{large}) and three deadline classes (\textit{strict}, \textit{moderate}, \textit{lenient}), producing \num{18} workflow instances per experiment.
\subsubsection{Text-to-Speech Censoring (T2SC)} transforms input text into speech while detecting and censoring profanities. It features an eight-function DAG with a mixture of parallelism and sequential audio-processing steps (Fig.~\ref{fig:WFs} (a)):
\begin{enumerate*}[(a)]
    \item\textit{GetInput} receives the input text and normalizes it for downstream processing;
    \item\textit{T2S} generates a raw speech waveform from the text;
    \item\textit{Conversion} transforms the audio to the target format (e.g., sample rate, codec);
    \item\textit{Compression} reduces the audio stream size before distribution;
    \item \textit{Profanity} runs in parallel with speech generation, detecting profane tokens in the input text;
    \item \textit{Merge} joins the compressed audio with the profanity annotations to build a time-aligned censoring map;
    \item \textit{Censor} applies muting or beep overlays at profanity locations;
    \item \textit{StoreAudio} persists the final censored audio file to the object store.
\end{enumerate*}
The parallel \textit{text2speech}/\textit{profanity} branch and subsequent merge–censor chain create a critical path dominated by processing-intensive audio processing, exposing \CL ability to coordinate concurrent branches and offload heavy functions under tight E2E deadlines.
\subsubsection{Regression Tuning (RT)}
\begin{figure}[!t]
\centering
\includegraphics[width=.9\columnwidth]{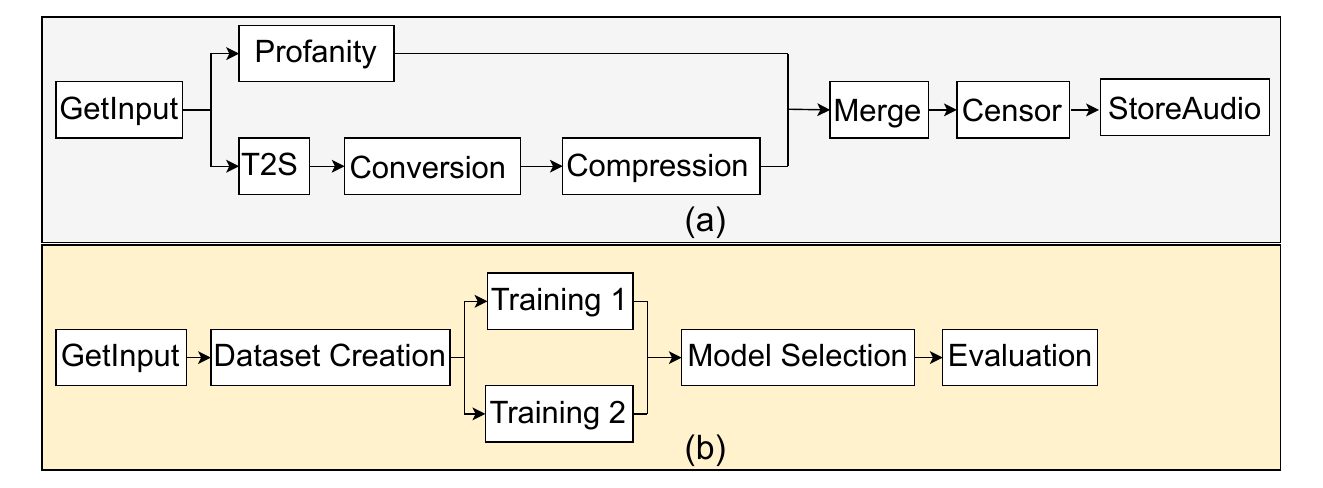}
\caption{\small{Case study serverless workflows.}} 
\label{fig:WFs}
\vspace{-15pt}
\end{figure}

performs E2E regression model selection on a structured dataset. The workflow follows a branched DAG of six functions (Fig.~\ref{fig:WFs} (b)):
\begin{enumerate*}[(a)]
    \item \textit{GetInput} receives the raw dataset and prepares it for processing;
    \item \textit{Dataset Creation} parses, cleans, and partitions the data;
    \item \textit{Training 1} and
    \item \textit{Training 2} train two regression models with distinct CPU and memory footprints;
    \item \textit{Model Selection} compares model accuracy and selects the superior model; 
    \item \textit{Evaluation} validates the chosen model on a held-out test set.
\end{enumerate*}
The parallel training stage creates a fork–join structure that stresses resource allocation, while the final selection–evaluation sequence introduces a delay-sensitive path. 
\subsection{Baseline methods}
Since no existing serverless workflow orchestration system jointly supports 
\begin{enumerate*}
\item explicit separation between intra- and inter-cluster orchestration and 
\item super-master–based coordination across federated clusters,
\end{enumerate*}
we compare \CL (CLU) against four baselines that reflect common design choices in serverless and multi-cluster workflow execution. 
\begin{figure*}[!t]
  \centering
  \begin{subfigure}[t]{0.52\textwidth}
    \centering
    \includegraphics[width=\linewidth]{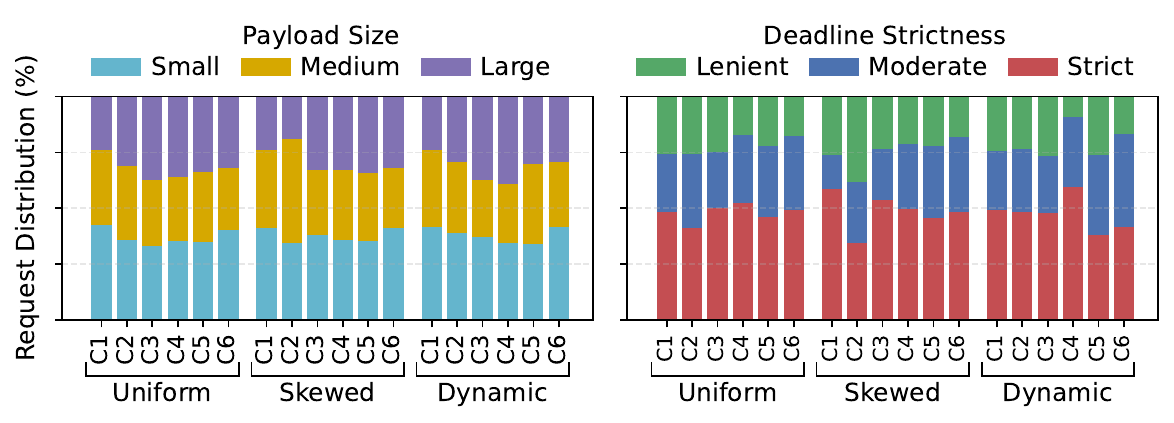}
    \caption{\small{Request size and deadline class distribution per cluster.}}
    \label{fig:req-size-deadline-combined}
  \end{subfigure}
  \hfill
  \begin{subfigure}[t]{0.46\textwidth}
    \centering
    \includegraphics[width=\linewidth]{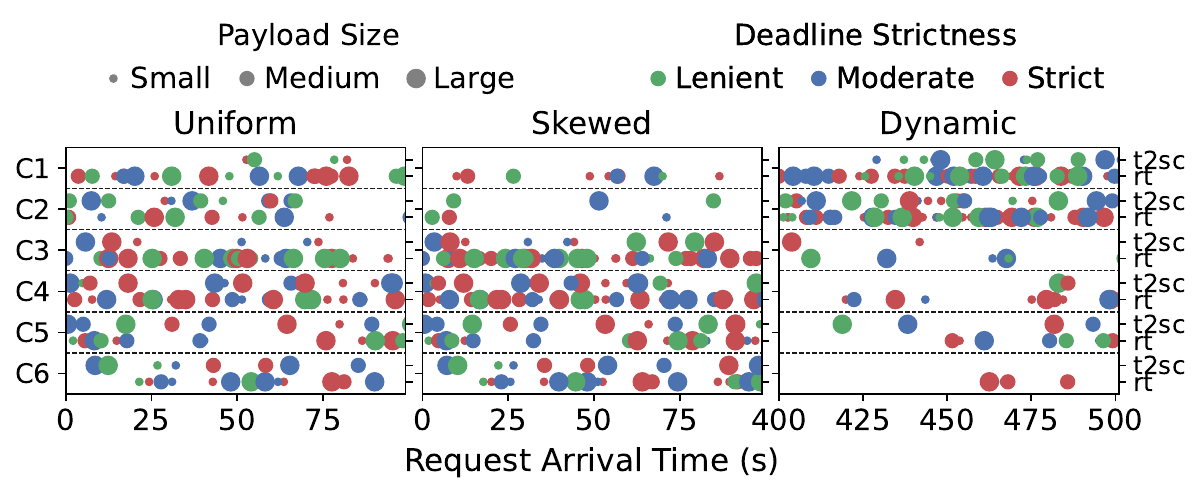}
    \caption{\small{Workflow arrival times per cluster.}}
    \label{fig:requests-arrivals_all}
  \end{subfigure}
  \caption{\small{Workload composition and temporal arrival behavior across clusters for different arrival rates.}}
  \label{fig:workload-characterization}
  \vspace{-10pt}
\end{figure*}
\subsubsection{NKS (Native K8s)} represents the standard K8s–Argo execution model, where each workflow function is deployed as a pod and orchestrated independently using default scoring and bin-packing policies. No workflow-level deadline awareness, dependency-aware start-time analysis, execution-mode selection, or inter-cluster orchestration is supported; all functions execute in the submission cluster. Autoscaling relies on the Horizontal Pod Autoscaler, which we configure to follow OpenFaaS-style scaling by comparing the number of active requests against a target concurrency per instance.
\subsubsection{CLI} isolates \CL \emph{intra-cluster orchestration}, enabling dependency-aware earliest-start analysis, resource and bandwidth feasibility checks, and execution-mode selection within a single cluster. However, inter-cluster orchestration is disabled, i.e., if a function cannot be placed locally in a deadline-feasible manner, it is forced to execute on the same cluster, potentially violating the deadline.
\subsubsection{RRX} extends CLI with deterministic inter-cluster offloading, assigning offloaded functions to clusters in a fixed round-robin order without deadline or load awareness.
\subsubsection{RNX} augments CLI with inter-cluster offloading by assigning each workflow category (type, payload size, deadline class) to a fixed randomly selected remote cluster. Like RRX, it ignores deadlines and load awareness.

\subsection{Experimental design}
We implemented all \CL components and algorithms in \texttt{Python 3.12}, interacting with the K8s API. 
\subsubsection{Orchestration parameters} 
We fix the control epoch to \mbox{$\Delta T=\qty{1}{s}$}, the heartbeat timeout to \mbox{$T_{\mathrm{fail}}=\qty{5}{s}$}, and the admissible load threshold to \mbox{$l=0.75$}.
\subsubsection{Concurrency limits and resource capacities}
For each worker $z_{ni}$, we set the concurrency capacity $R_{ni}$ proportional to its physical CPU core count, enforcing at most one non-preemptive function per core. The normalized cluster loads $\mathrm{L}_n(t)$ in Eq.~(\ref{eq:1}) and feasibility checks in Eqs.~(\ref{eq:12})--(\ref{eq:14}) use the instantaneous computational and bandwidth measurements exported by Prometheus/cAdvisor. Execution modes, warm execution, warm scaling, and cold scaling, are implemented through a custom autoscaling and container-lifecycle controller, designed to replicate \texttt{OpenFaaS}-style behavior while allowing explicit control over replica creation, cold-start delays, and concurrency limits on each worker node.
\subsubsection{Workflow instances}
We evaluate all size–deadline combinations defined by the workflow templates, resulting in \num{18} workload configurations. 
For the \emph{T2SC} workflow, input sizes are \textit{small} (\qty{500}{characters}), \textit{medium} (\qty{1250}{characters}), and \textit{large} (\qty{3750}{characters}), with deadline classes \textit{lenient}, \textit{moderate}, and \textit{strict} specified as $(130,180,150)$, $(100,130,180)$, and $(70,90,110)$, respectively, where each tuple corresponds to (small, medium, large). 
For the \emph{RT} workflow, dataset sizes are \textit{small} (\num{25000}), \textit{medium} (\num{100000}), and \textit{large} (\num{250000}) samples, combined with \textit{lenient} $(120,180,250)$, \textit{moderate} $(100,150,200)$, and \textit{strict} $(80,110,150)$ deadlines.
\subsubsection{Requests arrival model}
Workflow requests are generated by first selecting a concrete workflow instance and then assigning an arrival time. Workflow instances are drawn from a \emph{Zipf distribution}~\cite{cherkasova2004analysis} over the $K=18$ size–deadline combinations, with selection probability of $i^{th}$ instance \mbox{$\mathcal{P}(i)=\frac{1/i^{\alpha}}{\sum_{j=1}^{K}1/j^{\alpha}},\alpha=0.75$}. Workflow request arrivals follow a \emph{Poisson} process. We evaluate three arrival regimes by controlling the arrival rate $\lambda$ over time:
\begin{enumerate*}[a]
 \item \textit{uniform load} applies a fixed and identical arrival rate \mbox{$\lambda=0.33$} across all clusters;
 \item \textit{skewed load} assigns heterogeneous but fixed arrival rates (\mbox{$\lambda\in\{0.1,0.05,0.55,0.5,0.4,0.4\}$}) to have persistent spatial load imbalance; and 
 \item \textit{dynamic load}, models temporal variations where $\lambda$ evolves over time, starting from the uniform regime and transitioning to a high-load configuration at predefined time points. The maximum arrival rate is empirically measured and set to \num{2} based on the highest stable rate observed on the federated testbed.
\end{enumerate*}

\section{Evaluation Results}\label{sec:res}
This section compares \CL (\emph{CLU}) with baselines, reporting average results and standard deviations.
\subsection{Workload analysis}
Fig.~\ref{fig:workload-characterization} summarizes the workload composition and temporal arrival behavior across the six clusters. Fig.~\ref{fig:req-size-deadline-combined} shows that the request mix is intentionally controlled and consistent across clusters and arrival regimes: the proportions of payload sizes (small/medium/large) and deadline classes (strict/moderate/lenient) remain stable under different loads. This design isolates the impact of orchestration decisions, such as queueing, execution-mode selection, and offloading, from variations in workload difficulty.
Fig.~\ref{fig:requests-arrivals_all} depicts workflow arrival times per cluster. Under the \emph{uniform} regime, arrivals are evenly distributed across clusters over the shown early execution window, yielding balanced concurrency. The \emph{skewed} regime introduces persistent spatial imbalance, with a subset of clusters receiving a disproportionate share of arrivals. The \emph{dynamic} regime shows a later execution window ($t\in\![400,500]$\,s) where arrivals become bursty and temporally correlated across clusters, creating short periods of overlapping submissions. This spatial--temporal variability increases instantaneous contention at dependency-constrained stages and thus stresses inter-cluster coordination and deadline-feasible offloading under non-stationary demand.
\subsection{Load-aware super-master analysis}
\begin{figure}[t!]
    \centering
    \includegraphics[width=.9\linewidth]{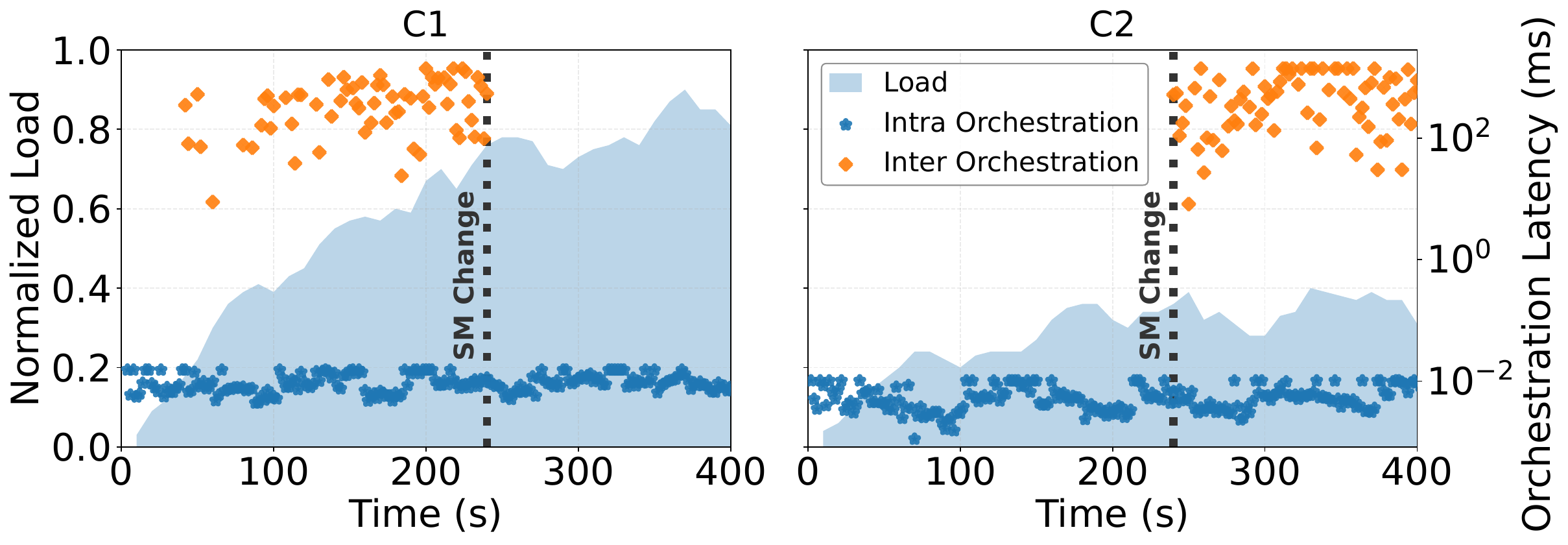}
    \caption{\small{Super-master behavior under increasing cluster load.}}
    \label{fig:sm-load}
\end{figure}

Fig.~\ref{fig:sm-load} traces the evolution of orchestration behavior over a \qty{400}{\second} interval as the load on cluster $C_1$ increases. The shaded area reports the normalized cluster load, while markers show the decision latency of \emph{intra-cluster} and \emph{inter-cluster} orchestration. In the initial phase ($t<$ vertical dashed line), the master of $C_1$ holds the super-master role and thus performs both local intra-cluster and inter-cluster orchestration. As the arrival rate to $C_1$ increases (from $0.4$ to $0.8$), its normalized load rises steadily, reflecting increasing local concurrency and queueing pressure. Throughout this phase, intra-cluster orchestration latency on $C_1$ remains low and stable, while inter-cluster orchestration incurs higher, but bounded, latency due to its global coordination scope.

When the load of $C_1$ reaches the admissible threshold (\mbox{$l=0.75$}; vertical dashed line), continuing inter-cluster orchestration on the same master would directly compete with local orchestration for control-plane resources. From this point onward, the super-master role is handled by the master of $C_2$, while $C_1$ continues exclusively with intra-cluster orchestration. Consequently, inter-cluster orchestration activity disappears from $C_1$, and its intra-cluster scheduling latency remains unchanged despite sustained workload. This behavior demonstrates that \CL confines global coordination to clusters with sufficient capacity headroom, preventing inter-cluster orchestration from amplifying contention and preserving stable intra-cluster orchestration under increasing load.
\subsection{Completion time analysis}
\begin{figure}[!t]
  \centering
  \includegraphics[width=.8\linewidth]{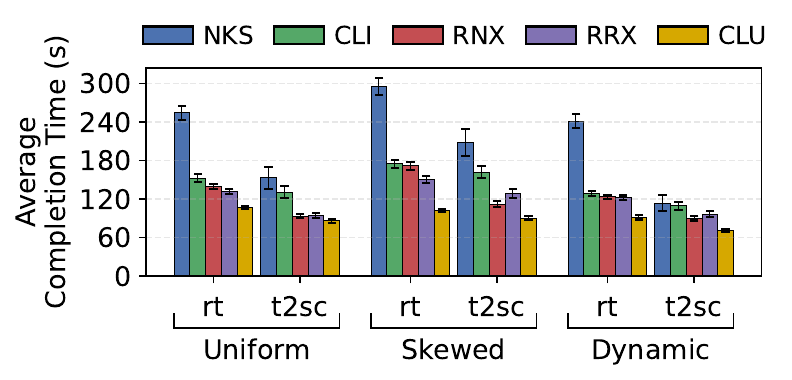}
  \caption{\small{Average workflow completion time across all clusters.}}
  \label{fig:completion-time-all}
  \vspace{-10pt}
\end{figure}

\begin{figure*}[t!]
  \centering
  \begin{subfigure}{.8\textwidth}
    \centering
    \includegraphics[width=0.4\linewidth]{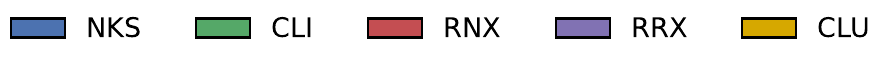}
  \end{subfigure}
  \vspace{0.4em}
  \begin{subfigure}{0.28\textwidth}
    \centering
    \includegraphics[width=\linewidth]{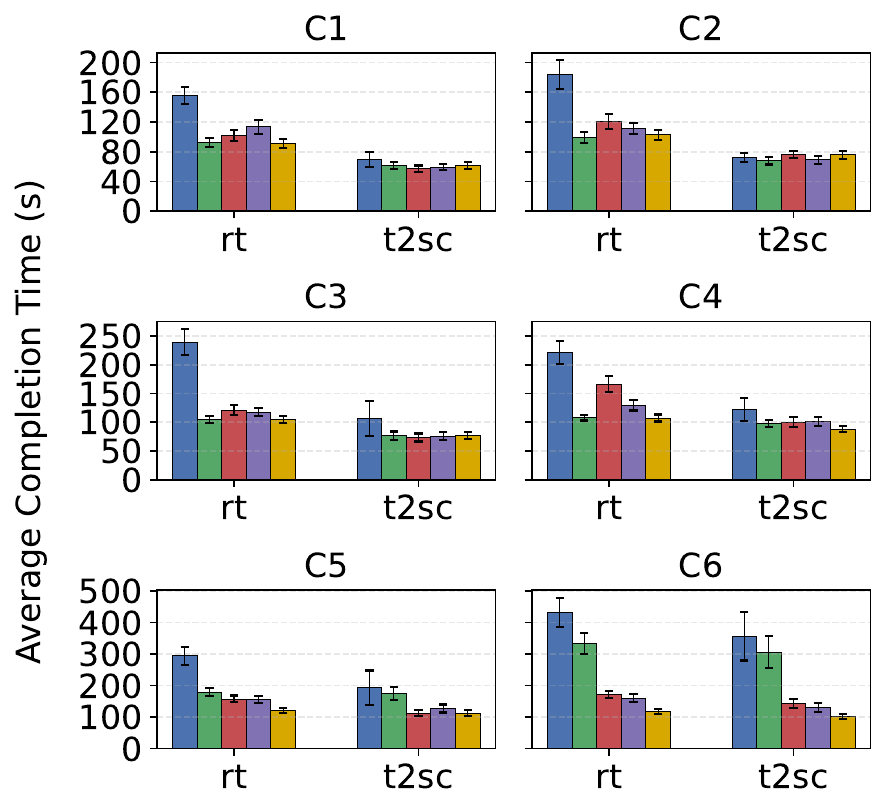}
    \caption{\small{Uniform.}}
    \label{fig:completion-lambda-0.05}
  \end{subfigure}
  \hfill
  \begin{subfigure}{0.28\textwidth}
    \centering
    \includegraphics[width=\linewidth]{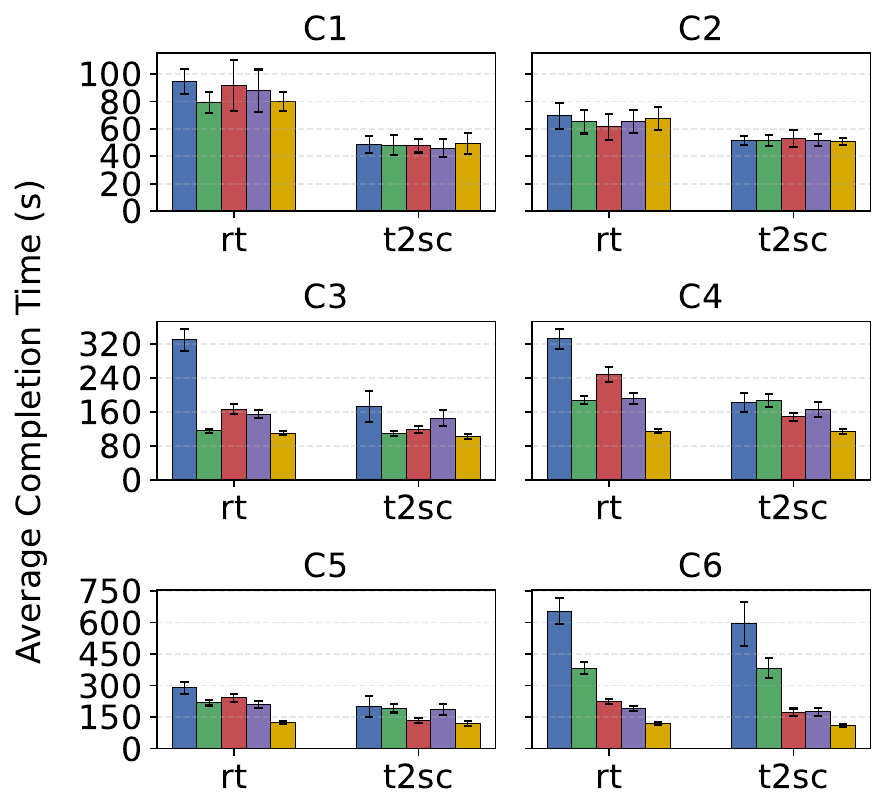}
    \caption{\small{Skewed.}}
    \label{fig:completion-lambda-0.1}
  \end{subfigure}
  \hfill
  \begin{subfigure}{0.28\textwidth}
    \centering
    \includegraphics[width=\linewidth]{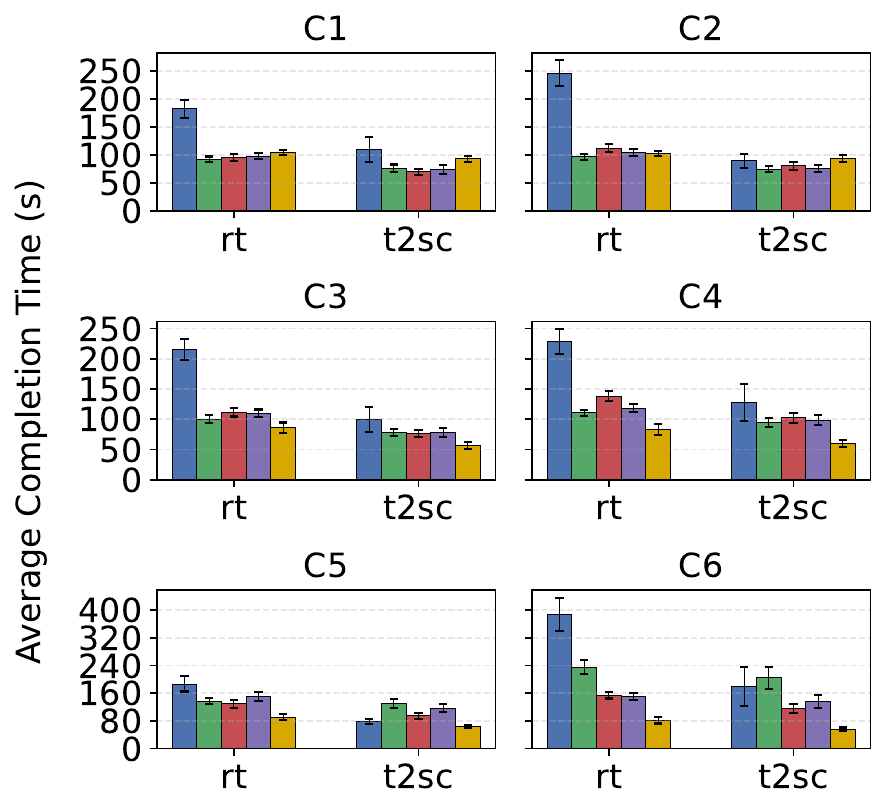}
    \caption{\small{Dynamic.}}
    \label{fig:completion-lambda-hybrid}
  \end{subfigure}
  \caption{\small{Average workflow completion time per cluster under different arrival regimes.}}
  \label{fig:completion-all-clusters}
  \vspace{-10pt}
\end{figure*}

Fig.~\ref{fig:completion-all-clusters} reports the average workflow completion time per cluster for RT and T2SC workflows under the three arrival regimes, while Fig.~\ref{fig:completion-time-all} summarizes the corresponding behavior across all clusters. Across all regimes, \emph{CLU} achieves the lowest (or tied-lowest) completion time and the smallest variability, indicating robust orchestration under both spatial and temporal load heterogeneity.
Under the \emph{uniform} load, arrivals are evenly distributed and completion times remain bounded. \emph{NKS} yields the highest completion times due to the lack of dependency-aware orchestration, execution-mode selection, and offloading, which amplifies queuing along workflow critical paths. \emph{CLI} improves over \emph{NKS} through mode-aware local orchestration but remains constrained under local saturation. \emph{RNX} and \emph{RRX} further reduce completion time by exporting load, yet uninformed target selection introduces unnecessary remote queueing. In contrast, \emph{CLU} achieves the lowest completion times by combining mode-aware intra-cluster execution with deadline-feasible inter-cluster placement, reducing RT completion time by \qty{15}{\percent}–\qty{21}{\percent} relative to \emph{RNX}/\emph{RRX} and by about \qty{10}{\percent} for T2SC (Fig.~\ref{fig:completion-time-all}). The per-cluster results (Fig.~\ref{fig:completion-all-clusters}) show that these gains are most pronounced on resource-constrained clusters ($C_5$–$C_6$).
\begin{figure*}[t!]
  \centering
  \begin{subfigure}[t]{0.5\textwidth}
    \centering
    \includegraphics[width=\linewidth]{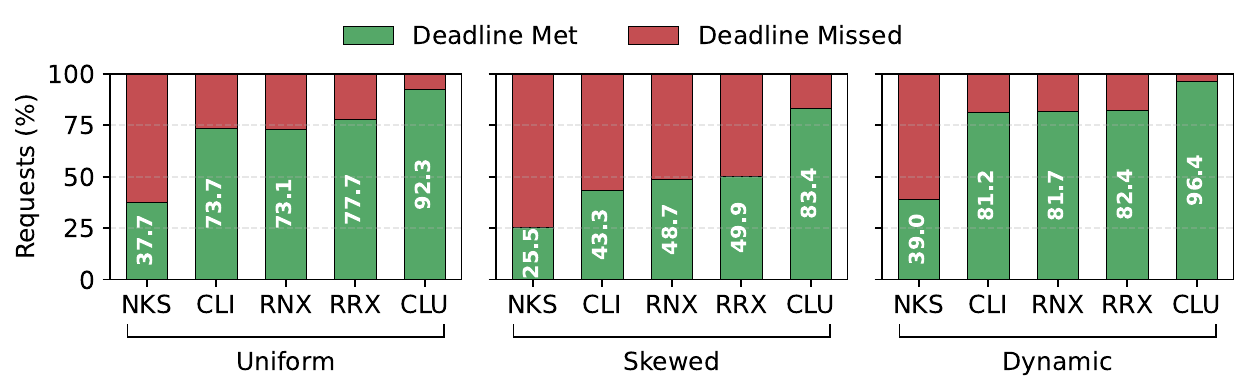}
    \caption{\small{Deadline satisfaction rate.}}
    \label{fig:deadline-violations-percentage}
  \end{subfigure}
  \hfill
  \begin{subfigure}[t]{0.48\textwidth}
    \centering
    \includegraphics[width=0.78\linewidth]{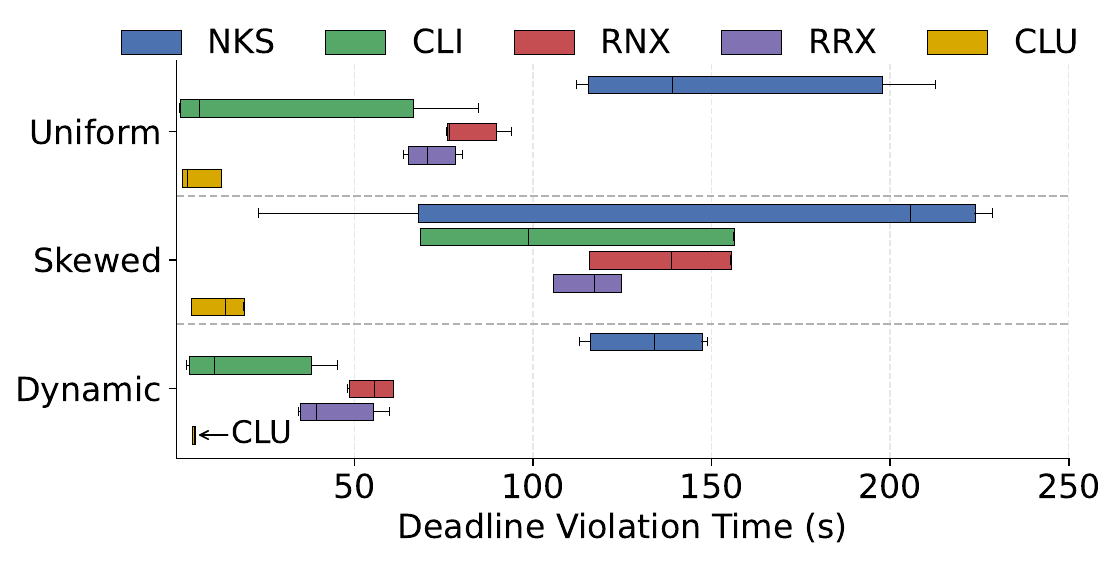}
    \caption{\small{Violation duration.}}
    \label{fig:deadline-violation-duration}
  \end{subfigure}
  \caption{\small{Deadline violation behavior across strategies and arrival rates.}}
  \label{fig:deadline-violation-summary}
  \vspace{-10pt}
\end{figure*}

Under the \emph{skewed} load with persistent spatial imbalance, performance gaps widen markedly. \emph{NKS} exhibits the highest completion times in hotspot clusters (e.g., $C_4$–$C_6$), while \emph{CLI} improves local execution but remains constrained by the lack of inter-cluster load redistribution. \emph{RNX} and \emph{RRX} partially alleviate local pressure, yet uninformed target selection results in elevated completion times. In contrast, \emph{CLU} maintains consistently lower RT completion times of about \qty{105}{\second} across clusters, yielding improvements of roughly \qty{30}{\percent} over \emph{RRX} and \qty{40}{\percent} over \emph{RNX}. Under temporal load variations (i.e., \emph{dynamic} regime), \emph{NKS}, \emph{RNX}, and \emph{RRX} exhibit elevated completion times during burst phases, while \emph{CLI} stabilizes local execution but remains affected by transient saturation. Fig.~\ref{fig:completion-time-all} shows that \emph{CLU} achieves the lowest completion times, improving RT by about \qty{21}{\percent}–\qty{24}{\percent} over \emph{RNX} and \emph{RRX} and T2SC by roughly \qty{22}{\percent}–\qty{30}{\percent}. The per-cluster results in Fig.~\ref{fig:completion-all-clusters} further show reduced variance across clusters, indicating that \emph{CLU} absorbs bursty arrivals through mode-aware intra-cluster orchestration combined with deadline-aware inter-cluster redistribution.
\subsection{Deadline violation analysis}
\begin{figure*}[t!]
  \centering
  \begin{subfigure}[t]{0.4\textwidth}
    \centering
    \includegraphics[width=\linewidth]{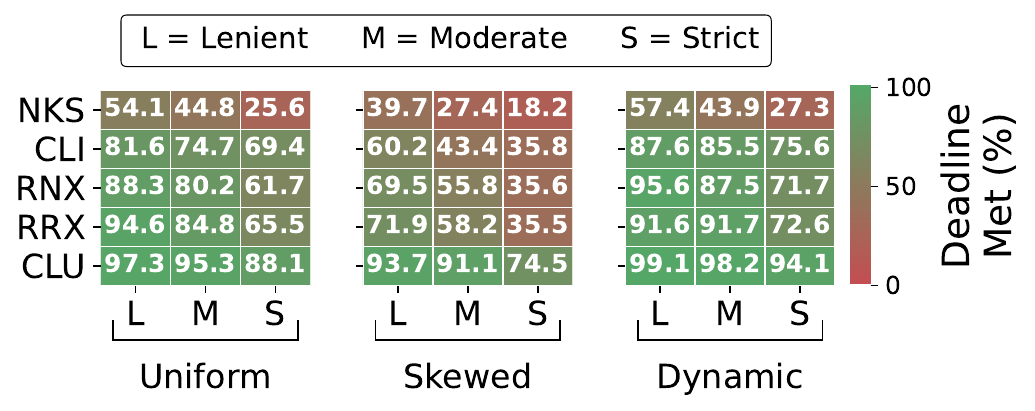}
    \caption{\small{Deadline satisfaction by deadline strictness.}}
    \label{fig:deadline-violation-deadline-types}
  \end{subfigure}
  \hfill
  \begin{subfigure}[t]{0.4\textwidth}
    \centering
    \includegraphics[width=\linewidth]{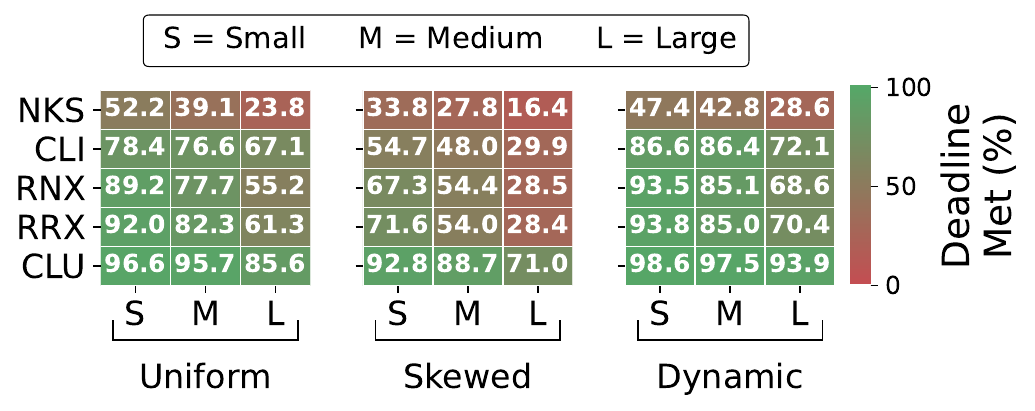}
    \caption{\small{Deadline satisfaction by payload size.}}
    \label{fig:deadline-violation-payload}
  \end{subfigure}
  \caption{\small{Deadline satisfaction across strategies and arrival rates under different workload constraints.}}
  \label{fig:deadline-violation-breakdown}
  \vspace{-5pt}
\end{figure*}

Fig.~\ref{fig:deadline-violations-percentage} reports deadline satisfaction rates across arrival regimes. Under the \emph{uniform} regime, \emph{CLU} achieves the highest satisfaction (\qty{92.3}{\percent}), outperforming \emph{RRX} (\qty{77.7}{\percent}), \emph{RNX} (\qty{73.1}{\percent}), and \emph{CLI} (\qty{73.7}{\percent}), while \emph{NKS} meets only \qty{37.7}{\percent} of deadlines. The gap reflects fundamental design differences, i.e., \emph{NKS} suffers from unbounded queueing, \emph{CLI} is constrained by local capacity, and \emph{RNX} and \emph{RRX} incur inefficient remote queueing due to uninformed offloading. In contrast, \emph{CLU} jointly optimizes execution mode and offloading decisions under deadline feasibility, preserving high satisfaction even at moderate load. Under the \emph{skewed} regime, persistent hotspots amplify these effects. As shown in Fig.~\ref{fig:deadline-violations-percentage}, satisfaction drops below \qty{50}{\percent} for \emph{CLI}, \emph{RNX}, and \emph{RRX}, whereas \emph{CLU} sustains \qty{83.4}{\percent}. Fig.~\ref{fig:deadline-violation-deadline-types} explains this gap; for \emph{strict} deadlines under \emph{skewed} regime, \emph{CLU} satisfies \qty{74.5}{\percent} of workflows, compared to \qty{35.8}{\percent} for \emph{CLI} and \qty{18.2}{\percent} for \emph{NKS}. Deadline-aware inter-cluster selection enables \emph{CLU} to relocate only those functions whose end-to-end completion can still meet constraints, preventing deadline loss in overloaded clusters. The \emph{dynamic} regime further stresses temporal adaptability. Fig.~\ref{fig:deadline-violations-percentage} shows that satisfaction for \emph{RNX} and \emph{RRX} falls below \qty{83}{\percent} during burst phases, while \emph{CLU} reaches \qty{96.4}{\percent}. 

Fig.~\ref{fig:deadline-violation-duration} quantifies violation severity. \emph{NKS} exhibits the longest violations, frequently exceeding \qty{200}{\second} under \emph{skewed} load due to cascading queue delays along workflow dependencies. \emph{CLI}, \emph{RNX}, and \emph{RRX} reduce violation duration but still incur delays on the order of tens of seconds. In contrast, \emph{CLU} consistently limits violation duration to single-digit seconds, indicating that even when deadlines are missed, violations remain tightly bounded. Finally, Fig.~\ref{fig:deadline-violation-payload} conditions satisfaction on payload size. Under \emph{skewed} load, large-payload workflows achieve below \qty{30}{\percent} satisfaction for all baselines, while \emph{CLU} maintains \qty{71}{\percent}, showing that \emph{CLU}’s deadline-aware policy explicitly accounts for both execution and inter-cluster transfer costs, which becomes critical as payload size increases.
\subsection{Offloading behavior and execution mode analysis}
\begin{figure*}[t!]
  \centering
  \begin{subfigure}[t]{0.38\textwidth}
    \centering
    \includegraphics[width=\linewidth]{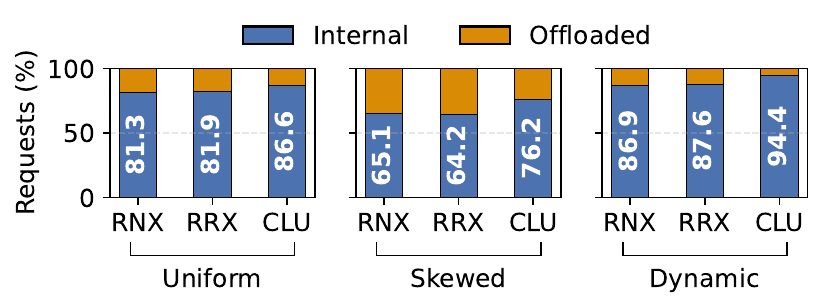}
    \caption{\small{Local vs. offloaded function execution.}}
    \label{fig:internal-external}
  \end{subfigure}
  \hfill
  \begin{subfigure}[t]{0.5\textwidth}
    \centering
    \includegraphics[width=\linewidth]{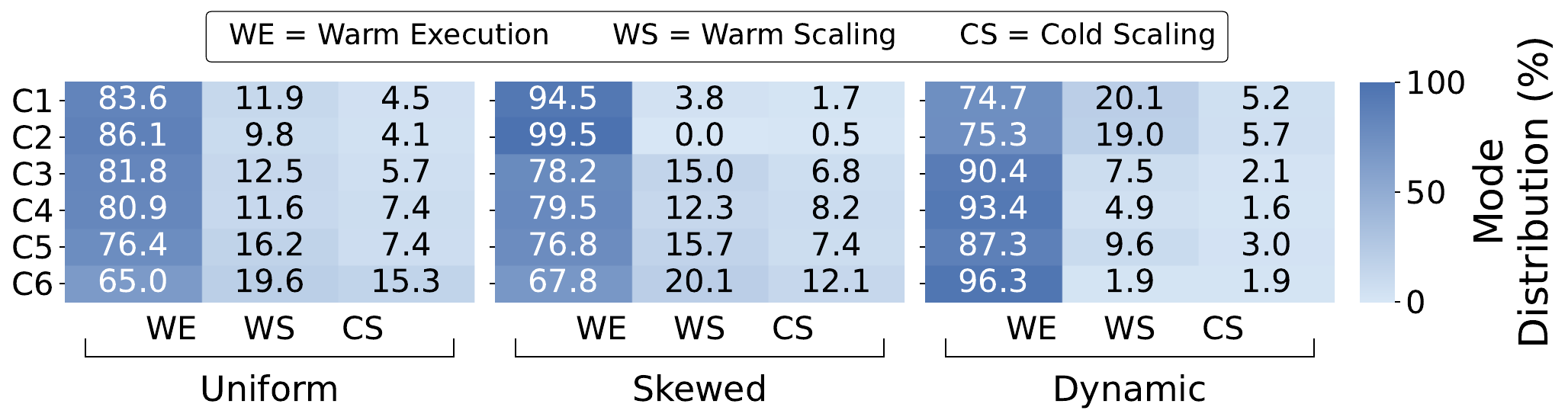}
    \caption{\small{Execution-mode distribution of \CL.}}
    \label{fig:mode-clu}
  \end{subfigure}
  \caption{\small{Offloading behavior and execution-mode selection under different arrival regimes.}}
  \label{fig:offloading-and-modes}
  \vspace{-10pt}
\end{figure*}

Fig.~\ref{fig:internal-external} quantifies how offloading-based strategies respond to increasing arrival pressure. Under the \emph{uniform} regime, most workflows are executed locally, but clear differences emerge: \emph{RNX} and \emph{RRX} execute about \qty{81}{\percent} of workflows internally, whereas \CL executes \qty{86.6}{\percent} locally, offloading selectively only when needed. Under the \emph{skewed} regime, offloading becomes essential. \emph{RNX} and \emph{RRX} offload aggressively yet inconsistently, retaining only \qty{65.1}{\percent} and \qty{64.2}{\percent} of workflows locally, respectively. In contrast, \CL maintains a higher internal execution share (\qty{76.2}{\percent}), indicating deadline-aware, targeted, and stable offloading decisions. Under the \emph{dynamic} regime, \CL further increases local execution to \qty{94.4}{\percent}, reflecting its ability to absorb bursts through coordinated execution rather than reactive spillover.
Fig.~\ref{fig:mode-clu} reports the execution-mode distribution of \CL across clusters and regimes. Warm execution remains dominant, ranging from \qty{65.0}{\percent} to \qty{86.1}{\percent} under the \emph{uniform} regime, from \qty{67.8}{\percent} to \qty{99.5}{\percent} under \emph{skewed} load, and from \qty{74.7}{\percent} to \qty{96.3}{\percent} under the \emph{dynamic} regime. Warm scaling stays bounded, peaking at \qty{19.6}{\percent} (\emph{uniform}), \qty{20.1}{\percent} (\emph{skewed}), and \qty{20.1}{\percent} (dynamic). Cold scaling starts are generally limited but become noticeable on the most constrained cluster: up to \qty{15.3}{\percent} in the \emph{uniform} regime and \qty{12.1}{\percent} in the \emph{skewed} regime (both on $C_6$), while remaining below \qty{5.7}{\percent} in the \emph{dynamic} regime. 
\begin{figure}[!t]
        \centering
        \includegraphics[width=.8\linewidth]{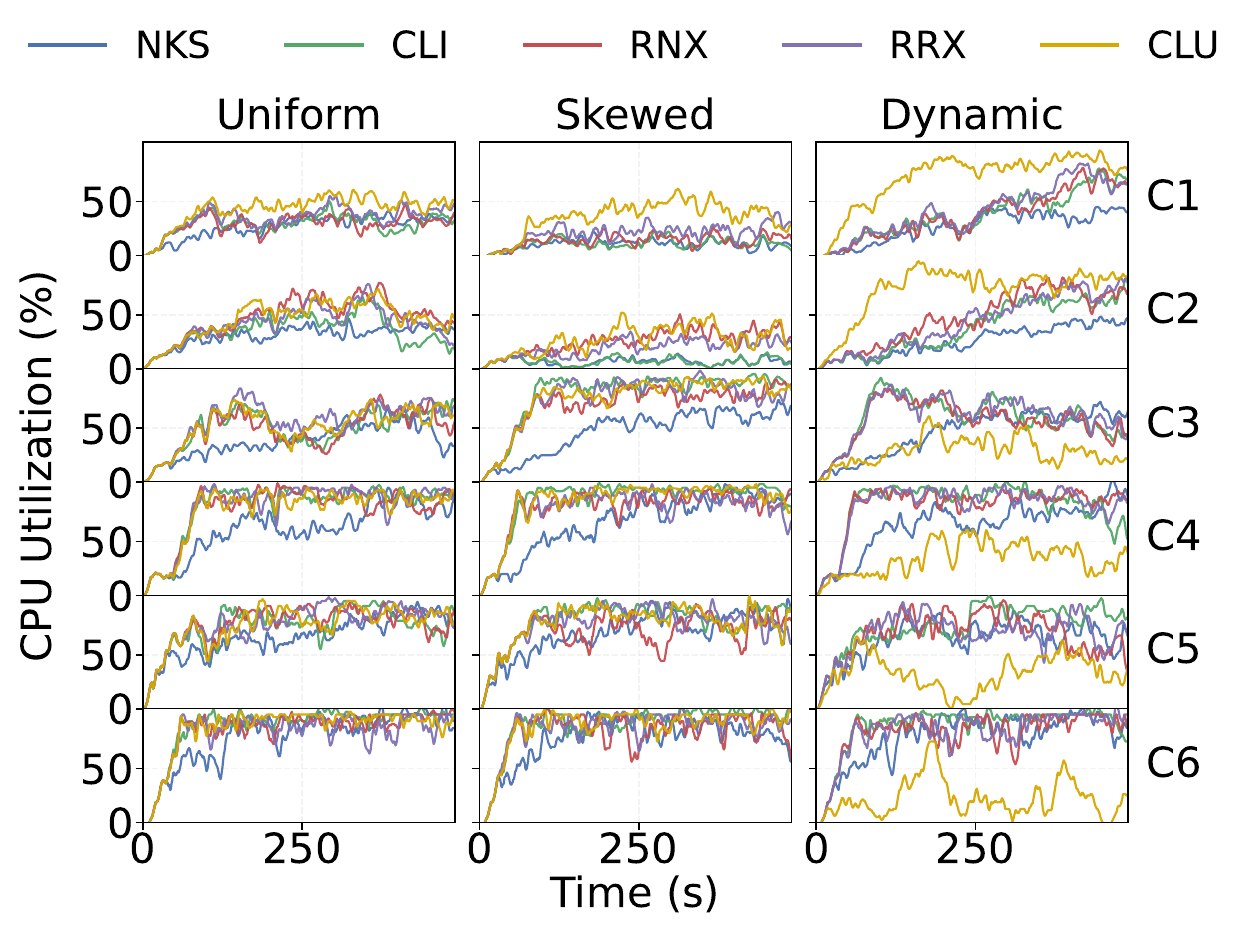}
        \caption{\small{Per-cluster CPU utilization in different methods.}}
        \label{fig:per-cluster-cpu-util}
    \vspace{-10pt}
\end{figure}
\subsection{CPU utilization analysis}
Fig.~\ref{fig:per-cluster-cpu-util} shows per-cluster CPU utilization over time under different arrival regimes. Under the \emph{uniform} and \emph{skewed} regimes, all strategies exhibit a similar ramp-up followed by steady operation on most clusters, indicating that resource usage is primarily driven by sustained workload intensity rather than orchestration choices.
Under the \emph{dynamic} load, clearer differences emerge. CPU utilization becomes uneven across clusters, reflecting the combined effect of bursty arrivals and inter-cluster execution decisions. \emph{CLU} shows higher utilization on some clusters (i.e., $C_1$–$C_2$) while maintaining noticeably lower utilization on others (i.e., $C_3$–$C_6$), indicating that execution pressure is redistributed across the federation rather than remaining locally concentrated. In contrast, \emph{NKS}, \emph{RNX}, and \emph{RRX} exhibit more uniformly elevated utilization across clusters, consistent with limited or uninformed load redistribution under \emph{dynamic} demand. 

\section{Conclusion}\label{sec:conclusion}
This paper presented \CL, a deadline-aware serverless workflow orchestration framework for federated multi-edge Kubernetes clusters. \CL combines mode-aware intra-cluster orchestration with super-master–based inter-cluster coordination, accounting for DAG dependencies, workflow deadlines, and heterogeneous compute and network conditions. We implemented \CL on six realistic edge clusters using OpenFaaS and Argo and evaluated it with two real workflows and \num{18} workload configurations. Results show that \CL reduces workflow completion time and deadline violations compared to four baselines. Future work will explore multi-objective and learning-based orchestration.

\balance
\bibliographystyle{ieeetr}
\bibliography{mainbib}
\end{document}